\documentclass[useAMS,usenatbib]{mnras}
\usepackage{graphicx}
\usepackage{amsmath}
\usepackage{amssymb}
\usepackage{color,bm}

\newcommand{\msun}{\ensuremath{{\rm M}_{\sun}}}

\include{defns}
\def\gsim{\;\rlap{\lower 2.5pt
 \hbox{$\sim$}}\raise 1.5pt\hbox{$>$}\;}
\def\lsim{\;\rlap{\lower 2.5pt
   \hbox{$\sim$}}\raise 1.5pt\hbox{$<$}\;}
\defcitealias{liu09a}{Liu et al. (2009a)}
\defcitealias{liu09b}{Liu et al (2009b)}

\def\del#1{{}}

\renewcommand{\vec}[1]{\mathbfit{#1}}
\newcommand{\ee}[1]{\times10^{#1}}

\newcommand{\pp}[2]{\frac{\upartial#1}{\upartial#2}}

\newcommand{\vv}{\bupsilon}
\newcommand{\vA}{\bupsilon_\rmn{A}}
\newcommand{\vs}{\bupsilon_\rmn{s}}
\newcommand{\vdi}{\bupsilon_\rmn{di}}
\newcommand{\vD}{\bupsilon_\rmn{D}}
\newcommand{\gas}{\mathrm{g}}
\newcommand{\CR}{\mathrm{c}}
\newcommand{\eps}{\varepsilon}

\title[Cosmic ray-driven galactic winds: streaming or diffusion?]{Cosmic ray-driven galactic winds: streaming or diffusion?}
\author[Wiener et al.]{Joshua Wiener$^{1,2}$, Christoph Pfrommer$^{3}$, \& S. Peng Oh$^{1}$ \\
$^{1}$Department of Physics; University of California; Santa Barbara, CA 93106, USA. \\
$^{2}$Department of Astronomy; University of Wisconsin; Madison, WI 53706, USA. \\
$^{3}$Heidelberg Institute for Theoretical Studies, Schloss-Wolfsbrunnenweg 35, 69118 Heidelberg, Germany.}

\begin{document}
\bibliographystyle{mnras}

\pagerange{000--000} \pubyear{0000}
\maketitle

\label{firstpage}

\begin{abstract}
  Cosmic rays (CRs) have recently re-emerged as attractive candidates for mediating feedback in galaxies because of their long cooling timescales. Simulations have shown that the momentum and energy deposited by CRs moving with respect to the ambient medium can drive galactic winds. However, simulations are hampered by our ignorance of the details of CR transport. Two key limits previously considered model CR transport as a purely diffusive process (with constant diffusion coefficient) and as an advective streaming process. With a series of GADGET simulations, we compare the results of these different assumptions. In idealised three-dimensional galaxy formation models, we show that these two cases result in significant differences for the galactic wind mass loss rates and star formation suppression in dwarf galaxies with halo masses $M\approx10^{10}\,\rmn{M}_\odot$: diffusive CR transport results in more than ten times larger mass loss rates compared to CR streaming models. We demonstrate that this is largely due to the excitation of Alfv{\'e}n waves during the CR streaming process that drains energy from the CR population to the thermal gas, which is subsequently radiated away. By contrast, CR diffusion conserves the CR energy in the absence of adiabatic changes and if CRs are efficiently scattered by Alfv{\'e}n waves that are propagating up the CR gradient. Moreover, because pressure gradients are preserved by CR streaming, but not diffusion, the two can have a significantly different dynamical evolution regardless of this energy exchange. In particular, the constant diffusion coefficients usually assumed can lead to unphysically high CR fluxes.
\end{abstract}


\section{Introduction}
\label{sec:intro}

What are the key physical ingredients responsible for galaxy formation?  This question is one of the most fascinating problems in modern astrophysics.  Recent cosmological hydrodynamic simulations have demonstrated the importance of momentum and energy feedback by star formation and supernovae to slow down star formation in galaxies, to obtain realistic rotation curves of disk galaxies, to match the small efficiency of converting baryons to stars in low-mass halos, and to move gas and metals out of galaxies into the intergalactic medium by means of galactic winds \citep{schaye10,guedes11,puchwein2013,hopkins14,marinacci14,vogelsberger14}. This progress of galaxy formation models is remarkable but it comes with a caveat. Feedback in these numerical calculations has thus far been implemented with simplified sub-grid prescriptions and modelled by phenomenological parameters that are tuned to reproduce global properties of galaxies such as the stellar mass fraction with halo mass, thus limiting the predictive power of these simulations. All approaches aim at avoiding the deposition of thermal energy in the dense phase of the interstellar medium, which would be very quickly radiated away, leaving behind little dynamical effect on the gas and producing a population of galaxies that is too compact in comparison to observations.

Three leading physical feedback mechanisms rely on momentum and energy deposition in the form of kinetic energy, ultra-violet (UV) radiation or via CRs. Momentum-driven winds form when radiation pressure acts efficiently on dust grains and atomic lines in dense gas and impart velocity kicks that expel the gas, provided it exceeds the escape velocity. While this idea has a long history, there has been some renewed interest due to its potential importance in galaxy formation \citep{odell67,chiao72,elmegreen83,ferrara93,scoville03,scoville01,thompson05,murray05,murray11,andrews11,zhang12}. While this mechanism yields a promising explanation for outflows in strong starburst galaxies, it is unknown whether the high opacities that are needed for radiation pressure to be efficient can be obtained in Milky-Way type galaxies, particularly at larger galactocentric radii \citep{krumholz12,krumholz13,rosdahl15,skinner15}. Moreover, this mechanism is not efficient at dwarf scales due to the smaller metallicities and UV photon production rate. Nonetheless, such systems still have very low star formation efficiencies, which challenge models which rely solely on radiation pressure feedback.

Another important source of free energy is CRs and magnetic fields, which are in rough equipartition with the thermal gas in our Galaxy \citep{boulares1990}. As CRs stream along field lines, they excite Alfv{\'e}n waves via the ``streaming instability'' \citep{kulsrud69}. CRs subsequently scatter off of these waves, which transfers momentum from the CRs to the gas. Furthermore, as the waves damp, net energy transfer can take place between CRs and gas. This momentum and energy transfer can drive a wind, as suggested by a number of theoretical works \citep{1975ApJ...196..107I, 1991A&A...245...79B, 1996A&A...311..113Z, 1997A&A...321..434P, 2002A&A...385..216B, 2008ApJ...687..202S, 2008ApJ...674..258E, 2010ApJ...711...13E, 2010MNRAS.402.2778S, 2012A&A...540A..77D}. By simultaneously solving the hydrodynamical equations and the CR transport equation along a one-dimensional (1D) flux tube under the action of self-generated diffusion and advection with the wind and the self-excited Alfv{\'e}n waves, \citet{recchia16} study implications of the launch and evolution of CR-driven winds on the CR spectrum.

Indeed, recent work has demonstrated that CR-mediated winds can expel gas from the disk and efficiently regulate star formation in three-dimensional (3D) hydrodynamic simulations of galaxy formation \citep{uhlig12, booth13, salem14a, salem14b, Pakmor2016,ruszkowski16} and local 3D simulations of the interstellar medium \citep{hanasz13, 2016ApJ...816L..19G, Simpson2016}.  CR-driven winds impart not only energy and momentum to the gas at the wind base, but can continuously repower the gas during its ascent because of dynamic and thermal coupling of CRs to the gas. An important feature is that CRs do not suffer radiative losses that a thermally driven wind would have, and so can potentially accelerate gas over much larger distances.

In this paper, we study the effects of different assumptions on CR transport and their ability to launch a wind. Recent numerical work has made a number of different assumptions regarding CR streaming (assuming it occurs at the sound speed, \citealt{uhlig12}; or ignoring it, \citealt{booth13,salem14a,Pakmor2016}) and diffusion (using a fixed isotropic diffusion coefficient as in \citealt{booth13,salem14a} or a fixed anisotropic diffusion coefficient, \citealt{Pakmor2016}). These assumptions strongly affect the CR profiles and their ability to drive a wind.

There is no complete consensus on dominant CR transport mechanisms. For instance, the diffusion coefficient depends on the nature of the dominant wave damping mechanism \citep{wiener13a}. Indeed, there are circumstances where CR scattering may be dominated by extrinsic turbulence rather than self-generated waves \citep{yan04,yan08}. On the other hand, for energies below 100 GeV - 1 TeV, self-generated turbulence due to propagating CRs could be dominating over externally driven turbulence in the galactic halo, potentially redering the assumption of an external diffusion coefficient to be unrealistic \citep[see e.g.,][]{aloisio15}. At the same time, it is important to understand how wind driving depends on the assumed CR transport mechanism, the likely validity of these assumptions, and situations where they could potentially lead us astray. For instance, the popular assumption of a constant diffusion coefficient is unlikely to be correct in detail. We note that we do not intend to accurately simulate all aspects of microscopic CR transport or CR driven winds in this work. We instead aim at demonstrating that wind properties (wind speed, mass loss rate, etc.) depend strongly on the type of CR transport assumed, and as such we must learn more about CR transport if we want to accurately model galactic winds.

This paper is structured as follows. In Section~\ref{sec:theory}, we briefly review canonical theory for CR transport in the self-confinement limit. In Section~\ref{sec:simulations}, we describe our GADGET simulation setup. In Section~\ref{sec:results}, we present results for 3D dwarf galaxy simulations, as well as 1D homogeneous tubes. We conclude in Section~\ref{sec:conclusions}.

\section{Cosmic-ray transport}
\label{sec:theory} 
\subsection{Equations}
\label{sec:theory_equations}
The evolution equations for a gas composed of thermal particles and CRs are a modified set of standard fluid equations, with an extra fluid equation for the CRs \citep[e.g.,][]{Guo08a,Pfrommer2016}. This system is supplemented by Poisson's equation and reads as follows,
\begin{eqnarray}
    \pp{\rho}{t}+\bm{\nabla\cdot}(\rho \vv )&\!\!=\!\!&0,
    \label{mass}\\
    \pp{(\rho\vv)}{t}+\bm{\nabla\cdot}(\rho\vv\vv^\rmn{T} + P)
    &\!\!=\!\!& -\rho\nabla\Phi,
    \label{momentum}\\
    \pp{\eps_\gas}{t}+\bm{\nabla\cdot}(\vv\eps_\gas)
    &\!\!=\!\!& -P_\gas\bm{\nabla\cdot}\vv+|\vA\bm{\cdot\nabla} P_\CR|\nonumber\\
    &&\hfill{}+\Gamma_\gas+\Lambda_\gas,
    \label{energy}\\
    \pp{\eps_\CR}{t}+\bm{\nabla\cdot}\vec{F}_\CR
    &\!\!=\!\!& -P_\CR\bm{\nabla\cdot}\vv-|\vA\bm{\cdot\nabla} P_\CR|\nonumber\\
    &&\hfill{}+\Gamma_\CR+\Lambda_\CR,
    \label{CRenergy}\\
    \vec{F}_\CR &\!\!=\!\!& \vv\eps_\CR + \vs (\eps_\CR+P_\CR)
    \hfill{}- \kappa\bm{\nabla} \eps_\CR,\nonumber\\
    \bm{\nabla}^2\Phi &\!\!=\!\!&4\upi G (\rho +\rho_\rmn{dm} + \rho_*).
    \label{Poisson}
\end{eqnarray}
Note that these equations neglect magnetic fields and assume isotropic CR (streaming and diffusion) transport relative to the gas rest frame.\footnote{While this represents a strong simplification to the correct description of anisotropic CR transport, we believe that this approach facilitates the clean separation of the effects of CR streaming and diffusion on the hydrodynamics without additional complications such as magnetic dynamo action or the possibility of a Parker instability.} Here, $\rho$, $\vv$, $P_\gas$, $\eps_\gas$ are the gas density, mean velocity, pressure and internal energy density, which are related via the equation of state, $P_\gas=(\gamma_\gas-1)\eps_\gas$, where $\gamma_\gas=5/3$. $P_\CR$ and $\eps_\CR$ are the CR pressure and energy density that are related by $P_\CR=(\gamma_\CR-1)\eps_\CR$, where $\gamma_\CR=\rmn{d}\ln P_\CR/\rmn{d}\ln\rho$ at fixed CR ``entropy''. The total pressure is denoted by $P=P_\gas+P_\CR$. The terms $\Gamma_i$ and $\Lambda_i$ ($i\in\{\gas,\CR\}$) are gain and loss terms, respectively.

$\vA=\vec{B}/\sqrt{4\upi\rho}$ is the Alfv{\'e}n speed, $\vs =-v_\rmn{A} \bm{\nabla} P_\CR/ |\bm{\nabla} P_\CR|$ is the isotropic Alfv{\'e}n streaming velocity down the CR gradient, and $\kappa$ is the CR diffusion coefficient. As a result, a population of CRs moves at bulk drift speed $\vD$ with respect to the rest frame of the ambient medium. $\vD = \vs + \vdi$ is composed of the CR streaming velocity $\vs$ and the isotropic CR diffusion velocity $\vdi=-\kappa \bm{\nabla} \eps_{\rmn{c}}/\eps_{\rmn{c}}$.\footnote{Note that there are conflicting statements in the literature about the term ``streaming velocity'': we use the term drift velocity $\vD$ to describe the bulk motion of CRs with respect to the background medium and streaming velocity $\vs$ to describe the advective motion of CRs with the Alfv{\'e}n wave frame in the direction opposite to the CR pressure gradient.}  Since we do not explicitly follow magnetic fields, we will adopt a model for the Alfv{\'e}n speed. $\Phi$ is the gravitational potential given by Poisson's equation~\eqref{Poisson}, $\rho_\rmn{dm}$ and $\rho_*$ are dark matter and stellar densities. In our 3D galaxy simulations, we approximate the dark matter distribution by a fixed external NFW potential rather than following the Vlasov equation by means of $N$-body methods.

Equations~\eqref{mass} and \eqref{momentum} represent mass and momentum conservation. The latter includes gravity and an extra term for the CR pressure gradient (which is contained in $P$).  Equations~\eqref{energy} and \eqref{CRenergy} describe the evolution of the energy densities of the thermal gas and CRs, respectively. They have the following terms on the right-hand side: adiabatic losses/gains, Alfv{\'e}n wave heating (for the gas) and cooling (for the CRs), energy gain processes (e.g., supernova energy) and cooling losses (radiative cooling for the gas; Coulomb and hadronic cooling for the CRs). Regardless of the source of Alfv{\'e}n wave damping, the CR heating rate is always $|\vA\bm{\cdot\nabla} P_\CR|$, as long as CRs are self-confined \citep{wiener13a, ruszkowski16}. This is because in the self-confinement picture CRs are advected with the wave frame and electric fields vanish there. Hence, CRs cannot experience an impulsive acceleration and can only scatter in pitch angle. Upon transforming to the rest frame of the gas there are electric fields associated with the propagating Alfv\'en waves representing time-varying magnetic fields effectively leading to the quoted energy transfer from the CRs to the gas.

$\vec{F}_\CR$ is the CR flux density that can be split into advection (with the gas), CR streaming and diffusion (relative to the gas rest frame). As will be detailed in Section~\ref{sec:simulations}, our numerical scheme follows the Lagrangian evolution of the momentum distribution function of CRs, $f(\vec{x},\vec{p},t)$ (assuming an isotropic power-law distribution in particle momentum), with a momentum dependent streaming speed $\vs(\vec{x},p,t)$ and a kinetic energy-weighted spatial diffusion coefficient $\kappa(\vec{x},t)$ that in general depend on position and time \citep{1991A&A...245...79B}. For simplicity and in order to make our results better comparable to previous work with similar set-ups \citep{booth13, hanasz13, salem14a, Pakmor2016}, we neglect these momentum dependencies.

\subsection{Streaming or diffusion?}
\label{sec:stream_diffuse}

The basis for the streaming term introduced in equation~\eqref{CRenergy} is derived in \cite{skilling71} and relies on the existence of the streaming instability \citep{kulsrud69}. A population of CRs moving at bulk drift speed $v_{\rm D}$ with respect to the ambient medium, along the background magnetic field, will generate magneto-hydrodynamic (MHD) Alfv\'en waves if $v_{\rm D}$ is greater than the local Alfv\'en speed $v_{\rm A}$. CRs will give up energy to these waves, inducing a wave growth rate that depends explicitly on $v_{\rm D}$. Equilibrium is determined by equating this growth rate with the wave damping rates of the background plasma. This is the self-confinement picture, where the transport of CRs is governed by scattering off of self-generated Alfv\'en waves.

The CR transport equations clearly imply advection at velocity $\vv+\vs$ (with $\vs =v_\rmn{A}\boldsymbol{n}$ and $\bm{n}$ is a unit vector pointing down the CR gradient), as well as a diffusion term. These fluid equations are derived from the original pitch-angle averaged CR equation for the full distribution function $f(\vec{x},\vec{p},t)$, derived in the limit of large wave-particle scattering \citep{skilling71}. Rapid scattering renders the CRs nearly isotropic in the rest frame of the wave; they thus advect with the wave. Since the fluid moves with velocity $\vv$ and the wave that is resonantly generated by CRs moves at velocity $\vs$ with respect to the fluid, the CRs move at $\vv+\vs$ with respect to the lab frame. The finite wave-particle scattering rate $\nu$ implies some slippage with respect to the wave frame; to second order in $\nu$, this is given by the diffusion term. \citet{skilling71} found that for the applications he was considering, the diffusion term was small, consistent with the truncation of the series expansion to second order. 

This assumption is generally true for the galactic wind applications we are considering. In the self-confinement picture, the effective drift speed $v_{\rm D}$ can be found by balancing the wave growth rate (which is proportional to the CR anisotropy $(v_{\rm D}/v_{\rm A} -1)$ in the wave frame) with the wave damping rate. Key wave damping mechanisms are ion-neutral damping (due to friction between neutrals---which are not tied to magnetic field lines---and ions), non-linear Landau damping (due to wave-particle interactions between thermal ions and the low-frequency beat wave formed by two interfering Alfv{\'e}n waves), and turbulent damping (the cascade of Alfv{\'e}n waves to smaller scales due to externally sourced MHD turbulence). For the ionised plasma in a galactic wind, the latter two are relevant. When non-linear Landau damping is dominant, the CR drift speed is \citep{wiener13a}:  
\begin{equation}
v_{\rm D} = v_{\rm A} \left( 1+  0.03 \frac{(n_{-3}^{\rm i})^{3/4} T_{\rm 6}^{1/4}}{B_{\rm \umu G}^{}L_{z,{\rm kpc}}^{1/2} (n^{\CR}_{-10})_{}^{1/2} }  \gamma^{(\alpha-3)/2} \right),
\label{eqn:vD_NLLD}
\end{equation}
where $T_{\rm 6}=T/10^{6} \, {\rm K}$, $B_{\rm \umu G}=B/1 \, \umu G$, $L_{z,{\rm kpc}}=L_{z}/1 \, {\rm kpc}$, $L_{z}\equiv P_{\rm c}/\nabla P_{\rm c}$ is the scale height of CRs, $n^{\rm i}_{-3}=n_{\rm i}/10^{-3} \, {\rm cm^{-3}}$, $n^{\CR}_{-10}=n^{\CR}(\gamma > 1)/10^{-10} \, {\rm cm^{-3}}$, $\gamma$ is the CR Lorentz factor, and $\alpha$ is the 3D slope of the CR distribution function (we have scaled to $\alpha=4.6$, its value in the Milky Way). Note that $n^{\CR} (> \gamma) = 10^{-10} \gamma^{-1.6} \, {\rm cm^{-3}}$ roughly corresponds to a CR energy density in equipartition with a $\sim \umu$G B-field. Due to a combination of a steep momentum spectrum and Coulomb losses at low energies, the CR energy density peaks at $\gamma \sim$~a few \citep[e.g.,][]{ensslin07}, so we should focus on CR streaming at these energies. Similarly, if turbulent damping dominates, 
\begin{equation}
v_{\rm D} = v_{\rm A} \left( 1+  0.06 \frac{(n_{-3}^{\rm i})^{7/4} \Lambda_{-22}^{1/2}\,\tilde{\eps}^{1/2}}{B_{\rm \umu G}^{} n^{\CR}_{-10}}  \gamma^{\alpha-3.5} \right),
\label{eqn:vD_turb}
\end{equation}
where $\Lambda_{-22} \equiv \Lambda(T)/(10^{-22} \, {\rm erg \, cm^{3} \, s^{-1}})$, $\Lambda(T)$ is the radiative cooling function, and $\tilde{\eps} = \rho v^{3}/(L n^{2} \Lambda(T))$ (where $L$ is the outer scale of turbulence, and $v$ is the velocity at this driving scale) is the ratio of the turbulent dissipation rate to the radiative cooling rate\footnote{Note that the total damping rate is the sum of all damping processes. In the limit where one dominates, the higher of these drift speeds should always be used.}. The last expression assumes strong turbulence; if turbulence is weak and sub-Alfv{\'e}nic, then $\tilde{\eps} = \rho v^{3} \mathcal{M}_{\rm A}/(L n^{2} \Lambda(T))$, where $\mathcal{M}_{\rm A}$ is the Alfv{\'e}n Mach number \citep{lazarian16}. Whilst the properties of turbulence can vary greatly, $\tilde{\eps} < 1$ should be regarded as a bound; if $\tilde{\eps} > 1$, then the heating due to turbulent dissipation exceeds the radiative cooling rate, and the wind is likely thermally rather than CR driven. Respecting this bound implies that streaming is unlikely to be super-Alfv{\'e}nic, particularly in low density halo gas\footnote{The situation can be different in the disk, where densities are higher and/or ion-neutral damping can be important.}. Thus, CRs are likely to be tightly locked to the wave frame with $v_{\rm D} \sim v_{\rm A}$. 

It is difficult to accurately capture streaming in numerical calculations. CRs can only stream down their gradient, which means that spurious numerical oscillations can develop near extrema, where the gradient flattens \citep{sharma10}. A frequent approximation is to ignore CR streaming and to set $\vv+\vs \rightarrow \vv$ in the CR equations, as well as set the diffusion coefficient $\kappa$ to a constant value (of order that measured in the Milky Way). A few comments are in order: (i) in terms of the equations we have discussed, this is tantamount to elevating the second order terms over the first order terms, which is not consistent with the assumptions of the series expansion. (ii) In the self-confinement picture, the form of the diffusion coefficient depends on plasma conditions and the dominant damping mechanism. Neither of the known damping mechanisms in ionised plasma results in diffusive behavior. One can show that \citep{wiener13a}: 
\begin{equation}
\nabla\cdot(\kappa \nabla  P_{\rm c}) \approx \nabla \cdot ({P}_{\rm c} \bm{n} (v_{\rm D}-v_{\rm A})), 
\end{equation}
where $\bm{n}$ is a unit vector pointing down the CR gradient. If we examine the expression for $(v_{\rm D}-v_{\rm A})$ for turbulent damping, we see that it is independent of $\nabla P_{\rm c}$. CR transport in this case is then equivalent to streaming at velocity $v_{\rm D}$ down the CR gradient. For non-linear Landau damping, $(v_{\rm D}-v_{\rm A}) \propto L_{z}^{-1/2} \propto (\nabla P_{\rm c})^{1/2}$, again distinct from diffusion where the flux is proportional to $\nabla P_{\rm c}$. (iii) Treating CRs in the purely diffusive approximation means that an energy transfer term $\vA \bm{\cdot \nabla} P_{\rm c}$, whereby CRs heat the gas, is ignored (see section \ref{sec:sim_overview}). One of the main advantages of CRs in calculations is that unlike thermal gas, it is not subject to radiative cooling. If it transfers its energy to the thermal gas over a scale height, this energy can be radiated, eliminating this advantage. We shall see that ignoring this term therefore overestimates the efficacy of CR driven winds. (iv) If magnetic fields are tangled at small scales, then CRs could still execute a small scale random walk (with ultimately diffusive behavior) by following the field lines, even if they stream at the Alfv\'en speed. Additionally, magnetic mirroring may become important. In this regard, the only effect would be a longer time scale for CRs to spread out - the energy coupling via wave heating will always be present if we suppose the CRs are self-confined. Cross-field diffusion is enhanced for small scaled tangled fields. The CR mean free path could thus be set by field line wandering. For this reason, it is worthwhile to examine the diffusive case. (v) Streaming and diffusion have different behaviors (see below). 

A true diffusion term contains two spatial derivatives. The general solution of the diffusion problem is a convolution of the initial conditions with a Gaussian (the Greens function). This is why any diffusive medium will smooth out in density over time -- the slope of any initial density profiles will become flatter, and it will do this at all points evenly. In contrast, an advection term only has one spatial derivative. Hence, the shape of a given profile is maintained in a static medium (in a moving medium, additional adiabatic changes arise). Streaming does not fit in any of these two categories. Because the direction of the velocity field depends on the direction of the density gradient, the profile will flatten out at extrema and necessarily introduces sharp kinks from smooth initial conditions, which are continuous but non-differentiable. These kinks propagate at the Alfv{\'e}n speed away from CR extrema. At scales larger than the Alfv{\'e}n crossing time from any CR extremum (maximum and minimum), the outward advection preserves the shape of the CR energy density. At the extrema, the nature of the flattening is independent of the magnitude of the gradient, and cannot be correctly described as a diffusion process.

In our approach, we do not follow magneto-hydrodynamics. Hence our isotropic transport scheme may overestimate the true anisotropic CR (streaming and diffusion) transport along the direction of the local magnetic field vector. However, this may only impact the solution at the very early stages before CR feedback will have changed the local orientation of the fields through a (Parker-like) buoyancy instability to align them (at least to some extent) with the local CR pressure gradient \citep{hanasz13, Pakmor2016}. Certainly, more work is needed to explore the detailed effects of anisotropic transport.

Ultimately, it is a false dichotomy to argue if CR streaming or diffusion is correct: both must be taking place. We now examine the nature of solutions assuming one or the other is dominant.

\section{Numerical simulations}
\label{sec:simulations}

\subsection{Overview of the simulation setups}
\label{sec:sim_overview} 

To demonstrate the dynamical difference between CR streaming and diffusion, we perform a number of different simulation setups. We start with a suite of simulations of two isolated, initially spherically symmetric halos of mass $10^9$ and $10^{10}\, \rmn{M}_\odot$ (Sect.~\ref{sec:galaxies}). After the gas is allowed to cool radiatively, it collapses to form a disk and starts to form stars according to a subgrid-scale model. This model also accounts for CR acceleration at supernova remnants and follows the detailed CR physics (see Sect.~\ref{sec:physics}).  In this work, we explore several variants of CR transport to isolate the dynamical impact of CR advection, diffusion, and streaming separately. We are especially interested in why and how CRs can drive galactic winds and how efficiently this impacts the star formation rate (SFR). To this end, we simulate six models with different variants of CR transport:
\begin{enumerate}
\item {\bf No CR model.} We follow the hydrodynamics without injecting CRs.
\item {\bf CR advection model.} CRs are only advected with the gas, i.e., $v_\rmn{s}=v_\rmn{A}=0, \kappa=0$ in equation~\eqref{CRenergy}.
\item {\bf CR diffusion model.} CRs are advected and allowed to diffuse but not to stream, i.e., $v_\rmn{s}=v_\rmn{A}=0$ in equation~\eqref{CRenergy}.
\item {\bf CR streaming model.} CRs can additionally stream but do not diffuse, i.e., $\kappa=0$ in equation~\eqref{CRenergy}.
\item {\bf CR diffusion model -- heating on.} Same as in (iii), but including the Alfv{\'e}n wave heating/cooling terms ($\vA\bm{\cdot\nabla} P_\CR$) in equations~\eqref{energy} and \eqref{CRenergy}.
\item {\bf CR streaming model -- heating off.} Same as in (iv), but artificially dropping the Alfv{\'e}n wave terms.
\end{enumerate}

The last two cases serve to examine separately the role of wave heating in galactic winds regardless of CR transport mechanism. The streaming model with wave heating turned off is purely artificial and has no physical basis. The diffusion model with heating turned on can be thought of as a case where diffusion and streaming are both occurring, but the diffusive transport dominates. Wave heating is therefore present, but the CR transport is effectively diffusive. This model should be contrasted with our pure diffusion model where streaming is completely neglected.

There are physically consistent models for both of these diffusion-dominated situations. First, we assume that we are in the self-confinement picture but the waves are highly damped. In this case both diffusion and streaming occur, but the diffusion dominates. The wave heating term is present and waves primarily travel down the CR gradient. This corresponds to model (v). In contrast, we imagine a situation with MHD waves traveling in all directions with roughly equal strength. This is no longer the self-confinement picture -- CRs are scattered equally both up and down their gradient, so there is no net energy transfer and the wave heating term is absent. Similarly there is no net bulk motion so the transport is necessarily diffusion-dominated. This is our pure diffusion model (iii).

We reiterate that the above simple models are not realistic but are merely limiting cases. Even CRs that are very well self-confined will exhibit some minimal level of diffusion, and only in a perfectly balanced situation as described above will streaming vanish completely. We use these models for simplicity -- a realistic treatment would require self-consistent calculation of MHD turbulence based on local damping mechanisms and is not necessary for the purpose of this work.

To isolate the physical effects of CR transport without the complications of other galaxy formation processes, we then simulate an idealised setup consisting of a uniform periodic slab of background gas (Sect.~\ref{sec:fluxtubes}). We superimpose onto this background a plane-parallel Gaussian initial CR pressure distribution. Gravity is turned off, but the hydrodynamic forces are accounted for, so the CR pressure will affect the background gas. The finite CR pressure gradients in the initial conditions thus set the gas in motion. In this setting, CRs are only allowed to cool adiabatically and through exciting Alfv{\'e}n waves (in the case of streaming), but we intentionally neglect hadronic or Coulomb losses. For these simplified setups, we perform two types of experiments: 1. we evolve the system using CR diffusion with a constant diffusion coefficient, and 2. we follow CR streaming at the Alfv\'en speed $v_\rmn{A}$ (i.e. the strong coupling limit).

\subsection{Simulated physics}
\label{sec:physics}

Our simulations were performed with an updated and extended version of the distributed-memory parallel TreeSPH code {\sc Gadget-2} \citep{springel05}. It employs a tree algorithm to compute gravitational forces and a variant of the smoothed particle hydrodynamics (SPH) algorithm to compute hydrodynamic forces. The code conserves gas energy and entropy where appropriate, i.e., outside of shocked regions \citep{springel02}.

Our galaxy formation simulations account for radiative cooling assuming an optically thin gas of primordial composition in collisional ionisation equilibrium. We account for star formation according to a sub-resolution model by \citet{springel03} which assumes that star forming regions establish a self-regulated regime that is described by an effective equation of state. We adopt a characteristic time-scale of star formation of $t_{\star}=1.5$ Gyr, and a density threshold of star formation of $\rho_0\simeq0.1\,\mathrm{cm^{-3}}$ (in units of hydrogen atoms per unit volume). These parameters have been chosen so that the simulations reproduce the observed Schmidt-Kennicut law, which is mostly sensitive to the effective equation of state and robust to feedback prescriptions \citep{springel03}. 

We model CR physics with a formalism that accounts for the relevant CR gain and loss processes self-consistently while accounting for the CR pressure in the equations of motion \citep{pfrommer06, ensslin07, jubelgas08}. In our methodology, we follow the evolution of the non-thermal CR distribution in momentum space along the Lagrangian trajectories of each (discretized) gaseous fluid element. The CR distribution function is modelled by a (time- and space-dependent) power-law spectrum in CR momentum and is characterised by an amplitude, a low-momentum cut-off, and a fixed slope of the 1D distribution function, $\alpha^{(1)} = 2.5$.\footnote{Here, we use the effective 1D CR spectrum $f^{(1)} (p) \equiv 4\upi p^2 f(p)$.}  We include adiabatic CR transport processes (compression and rarefaction), and a number of physical source and sink terms, which modify the CR pressure of each particle. We consider CR acceleration by supernova remnants with a spectral index for the directly accelerated CR population of $\alpha_{\mathrm{SN}}^{(1)}=2.5$ and an acceleration efficiency (fraction of supernova feedback energy directed into the CR population) of $\zeta_{\mathrm{SN}}=0.3$. As CR loss processes, we include thermalization by Coulomb interactions and catastrophic losses by hadronic interactions, which are dominant at low and high CR momenta ($p\ll m_{\rmn{p}} c$ and $p\gg m_{\rmn{p}}c$), respectively.  In the absence of freshly injected CRs, Coulomb cooling increases the spectral break and hadronic CR interactions lowers the break. As a result, the CR population evolves towards an attractor solution that resembles the equilibrium distribution between injection and cooling \citep[see][for further details]{ensslin07, jubelgas08}.

In addition to CR advection, the code can follow CR diffusion and streaming \citep{uhlig12}, using a split-operator, explicit finite difference scheme in both cases. We validate our code against exact solutions in Appendix~\ref{sec:tests}. We assume equipartition between gas and magnetic pressure, $P_\gas\approx P_B$, which is observed in our Galaxy and other normal spiral galaxies \citep{thompson06}. This implies that the sound speed is a good proxy of Alfv{\'e}n speed, $v_\rmn{A}\approx c_\rmn{s}$ in the warm-hot interstellar medium (ISM). In general, $v_{\rm A}/c_{\rm s} \approx \beta^{-1/2} = (P_B/P_\gas)^{1/2}$. This similarity should even hold for several scale heights in an adiabatically expanding wind. 

\begin{table*}
\caption{Properties of the different CR transport model simulations.}
 \begin{center}
  \begin{tabular}{l l l c c c c c}
    \hline
    $M_{200}$ & CR Physics & Wave Heating & $v_{200}^a$ & max $v_z$ (disk)$^b$ & max $v_z$ (halo)$^c$ & $\Delta M_\rmn{tot}/M_0$ (disk)$^d$ & $\Delta M_\rmn{tot}/M_0$ (halo)$^e$ \\
    \phantom{\Big|}&  &  & [km s$^{-1}$] &  [km s$^{-1}$] & [km s$^{-1}$] & & \\
    \hline
    \hline
    10$^9\ \rmn{M}_\odot$ & No CRs & No & 16.3 & 2.6 & ~~7.7 & $+1.06$ & $-0.04$ \\
    & Advection   & No                  &     & 5.8 & ~~7.1 & $+0.89$ & $-0.05$ \\
    & Diffusion   & No                  &     & 4.1 &  31.0 & $-0.81$ & $-0.92$ \\
    & Streaming   & Yes                 &     & 3.8 &  12.5 & $+0.27$ & $-0.26$ \\
    & Diffusion   & Yes (alternate)     &     & 4.2 &  25.6 & $-0.83$ & $-0.93$ \\
    & Streaming   & No (alternate)     &     & 4.2 &  24.3 & $-0.50$ & $-0.69$ \\
    \hline
    10$^{10}\ \rmn{M}_\odot$ & No CRs & No & 35.1 & ~7.8  & 12.5 & $+1.76$ & $+0.13$ \\   
    & Advection     & No                  &      &  18.2 & 22.9 & $+1.70$ & $+0.13$ \\
    & Diffusion     & No                  &      &  22.6 & 49.7 & $-0.68$ & $-0.85$ \\
    & Streaming     & Yes                 &      & ~8.8  & 32.4 & $+1.58$ & $+0.10$ \\
    & Diffusion     & Yes (alternate)     &      & ~8.1  & 39.6 & $+0.53$ & $-0.25$ \\
    & Streaming     & No (alternate)      &      & ~5.0  & 57.2 & $-0.07$ & $-0.56$ \\
    \hline
    \label{tab:models}
  \end{tabular}
  \begin{quote}
    Notes:\\
    $^a$The virial velocity of the halo is defined here as $v_{200}=\sqrt{GM_{200}/R_{200}}$.\\
    $^b$Maximum vertical {\em disk} velocity: we average over the highest 100 values of $v_z$ at 1 Gyr within the top half of a central disk (10$^9\ M_\odot$ halos: $r_{\rmn{disk}}<5$~kpc and $0<z<1$~kpc; 10$^{10}\ M_\odot$ halos: $r_{\rmn{disk}}<10$ kpc and $0<z<2$ kpc, where $r_{\rmn{disk}}$ denotes the disk radius).\\
    $^c$Maximum vertical {\em halo} velocity: we average over the highest 100 values of $v_z$ at 1 Gyr within the intersection of the upper half-sphere of radius $R_{200}$ ($16.2$ and $34.9$ kpc) and a cylinder of radius $r_{\rmn{disk}}$ (as defined in $^b$).\\
    $^d$Fractional change in baryonic mass of the {\em disk} region (as defined in $^b$, including the part below the midplane) from 0 to 3 Gyr.\\
    $^e$Fractional change in baryonic mass of the {\em halo} region (as defined in $^c$, including the part below the midplane) from 0 to 3 Gyr.
  \end{quote}
 \end{center}
\end{table*}

Measurements of halo magnetic fields indicate $B \sim \umu$G fields \citep{haverkorn15}, i.e. comparable to disk $B$-fields. Interestingly, given that thermal pressure decreases from the disk to the halo, this suggests $\beta < 1$, i.e. $v_{\rm A} > c_{\rm s}$, and that adopting $v_{\rm A} \sim c_{\rm s}$ is reasonable or even conservative. Note also that in multi-phase gas, the $B$-field strength is observed to be independent of density for $n \leq 300 \, {\rm cm^{-3}}$ \citep{crutcher10}, and thus $\beta \sim 1, \ v_{\rm A} \sim c_{\rm s}$ in both warm photo-ionised gas and hot coronal gas. Of course, $v_{\rm A}$ (and $c_{\rm s}$) drop sharply in the cold gas; we explore the consequences of this in Wiener et al. (2016, in preparation).

In all of the simulations described here, the streaming time constraint is the same as described in \cite{uhlig12},
\begin{equation}
\Delta t<\epsilon\frac{1}{\lambda_\rmn{s} c_s}\left(\frac{m}{\rho}\right)^{1/3},
\end{equation}
where $\epsilon=0.004$, $\lambda_\rmn{s}=1$, $m$ and $\rho$ are the mass and mass density of an SPH particle under consideration. This criterion scales the timestep to the time it takes for a CR to stream across the mean interparticle spacing $(m/\rho)^{1/3}$ at speed $\lambda_\rmn{s} c_s$. We demonstrate numerical convergence for these parameter choices in Appendix~\ref{sec:res}. For simulations evolving with diffusion we use a constant diffusion coefficient of $\kappa=3\ee{28}$ cm$^2$ s$^{-1}$.

\begin{figure*}
\includegraphics{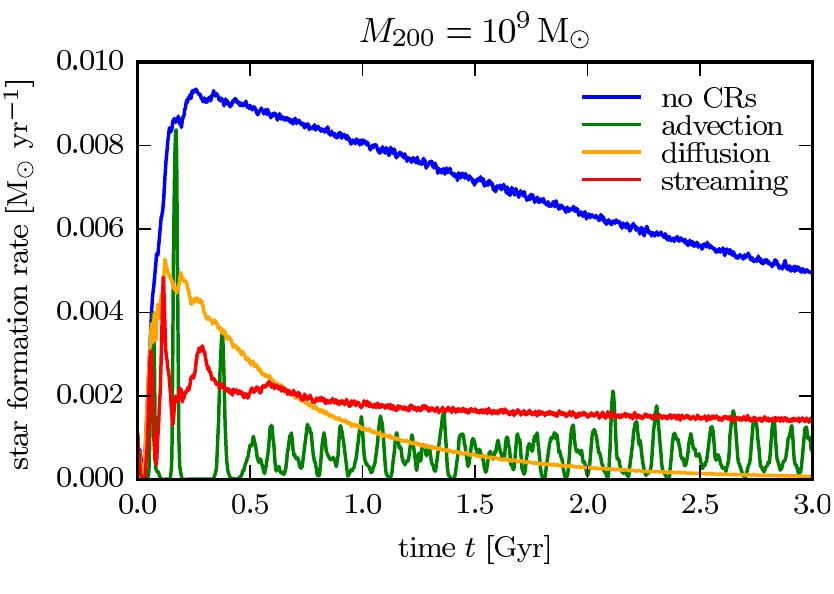}
\includegraphics{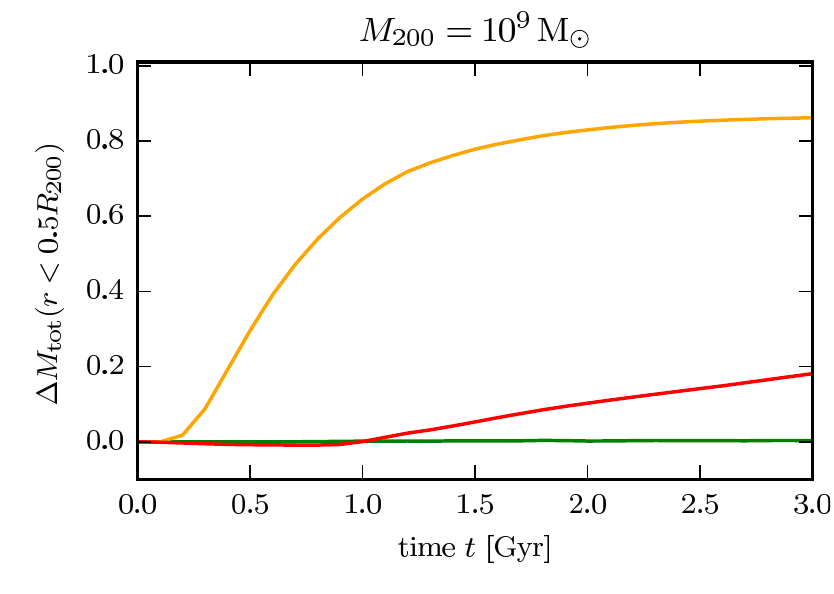}
\includegraphics{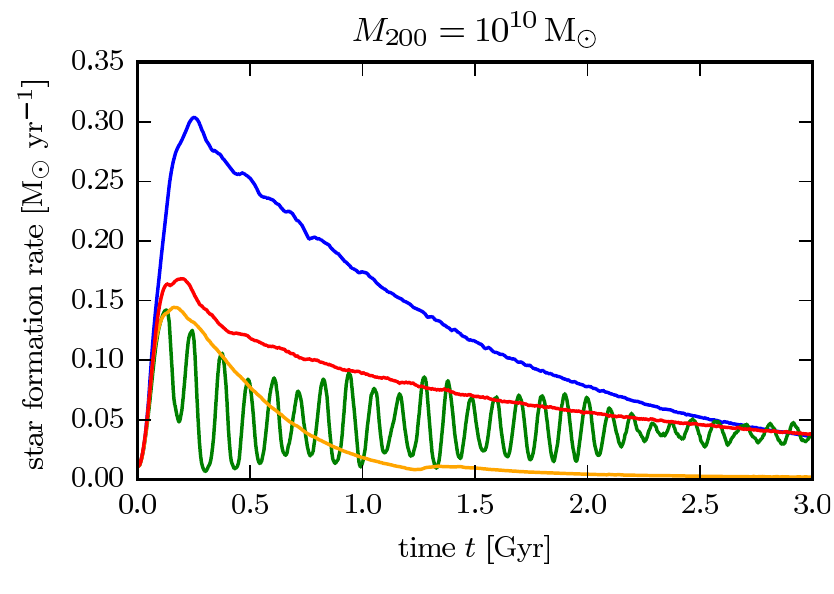}
\includegraphics{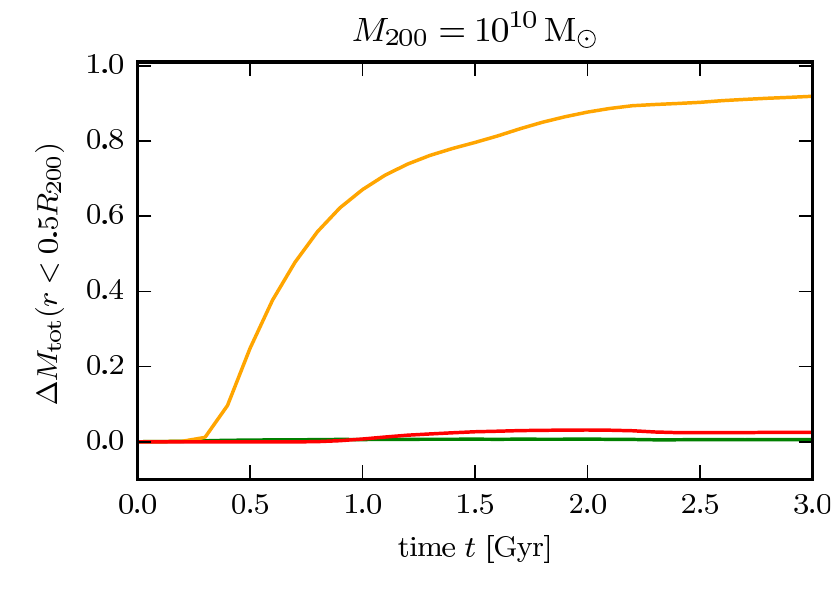}
\caption{Star formation rates (SFRs, left) and mass loss relative to simulations without CRs (right) as a function of time for the four feedback models. We compare simulations of halo masses 10$^{9}\ \rmn{M}_\odot$ (top) and  10$^{10}\ \rmn{M}_\odot$ (bottom). All CR feedback models suppress the SFR, with a larger suppression factor in the smaller galaxy. CR diffusion drives a strong outflow in both halos whereas CR streaming only causes a significant mass loss in the 10$^{9}\ \rmn{M}_\odot$ halo. }
\label{SFRplot}
\end{figure*}

\begin{figure*}
\includegraphics{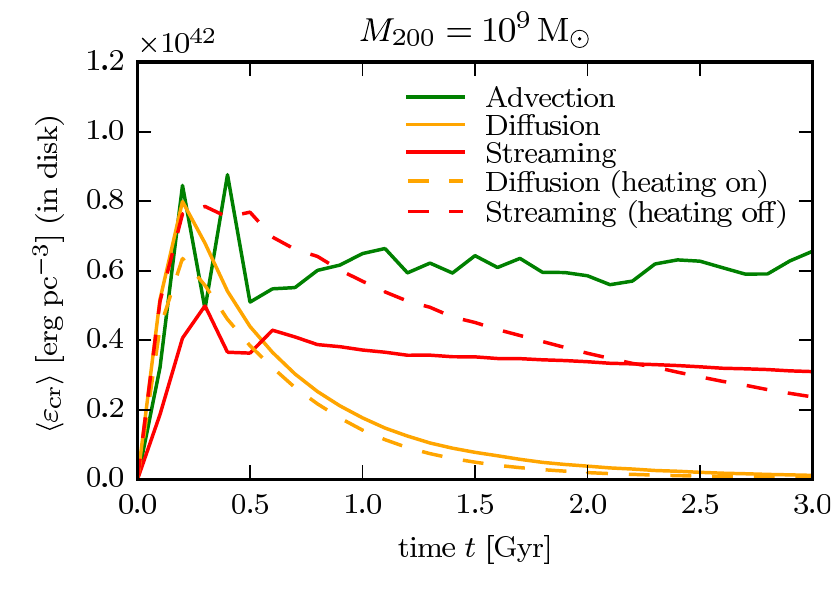}
\includegraphics{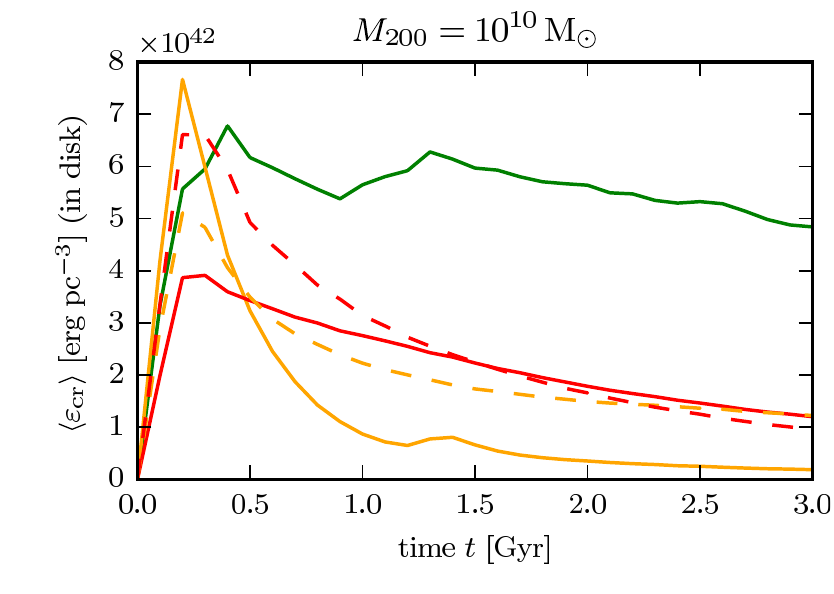}
\caption{Time evolution of the volume-averaged CR energy density $\langle\varepsilon_\rmn{cr}\rangle$ in a cylindrical disk of height $|z|<z_\rmn{disk}$ and radius $r_{\rmn{disk}}$ for the different CR models.  We compare simulations of halo masses 10$^{9}\ \rmn{M}_\odot$ (left; $z_{\rmn{disk}}=1$~kpc and $r_{\rmn{disk}}=5$~kpc) and 10$^{10}\ \rmn{M}_\odot$ (right; $z_{\rmn{disk}}=2$~kpc and $r_{\rmn{disk}}=10$~kpc).}
\label{ecr_bar}
\end{figure*}

\begin{figure*}
\includegraphics{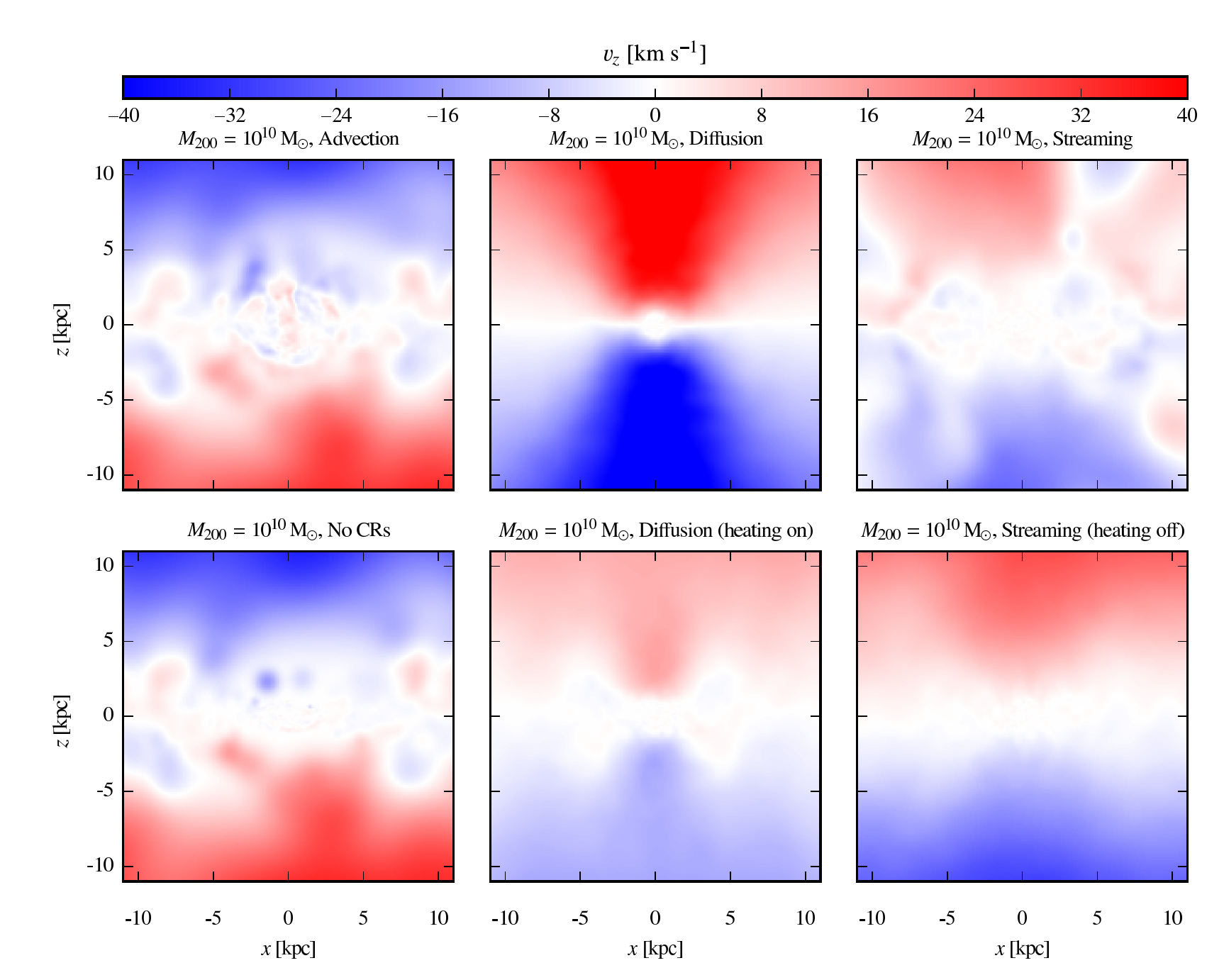}
\caption{Edge-on slices of vertical velocity $v_z$ through the centre of the $10^{10}\ \rmn{M}_\odot$ halo. Shown are the six different simulation models after 1 Gyr. Only models with active CR transport (i.e., streaming or diffusion) drive outflows from the disk (middle and right-hand panels). The outflow in the pure diffusion model is much stronger than in the streaming model due to the CR energy loss as a result of the Alfv{\'e}n wave cooling term, $\vA\bm{\cdot\nabla} P_\CR$. Artificially suppressing this term (bottom right panel) results in a stronger and faster outflow. Only models without the Alfv{\'e}n wave cooling term show vertical velocities that exceed the virial velocity $v_{200}=\sqrt{GM_{200}/R_{200}}=35~\rmn{km~s}^{-1}$ substantially.}
\label{vel10}
\end{figure*}

\begin{figure*}
\includegraphics{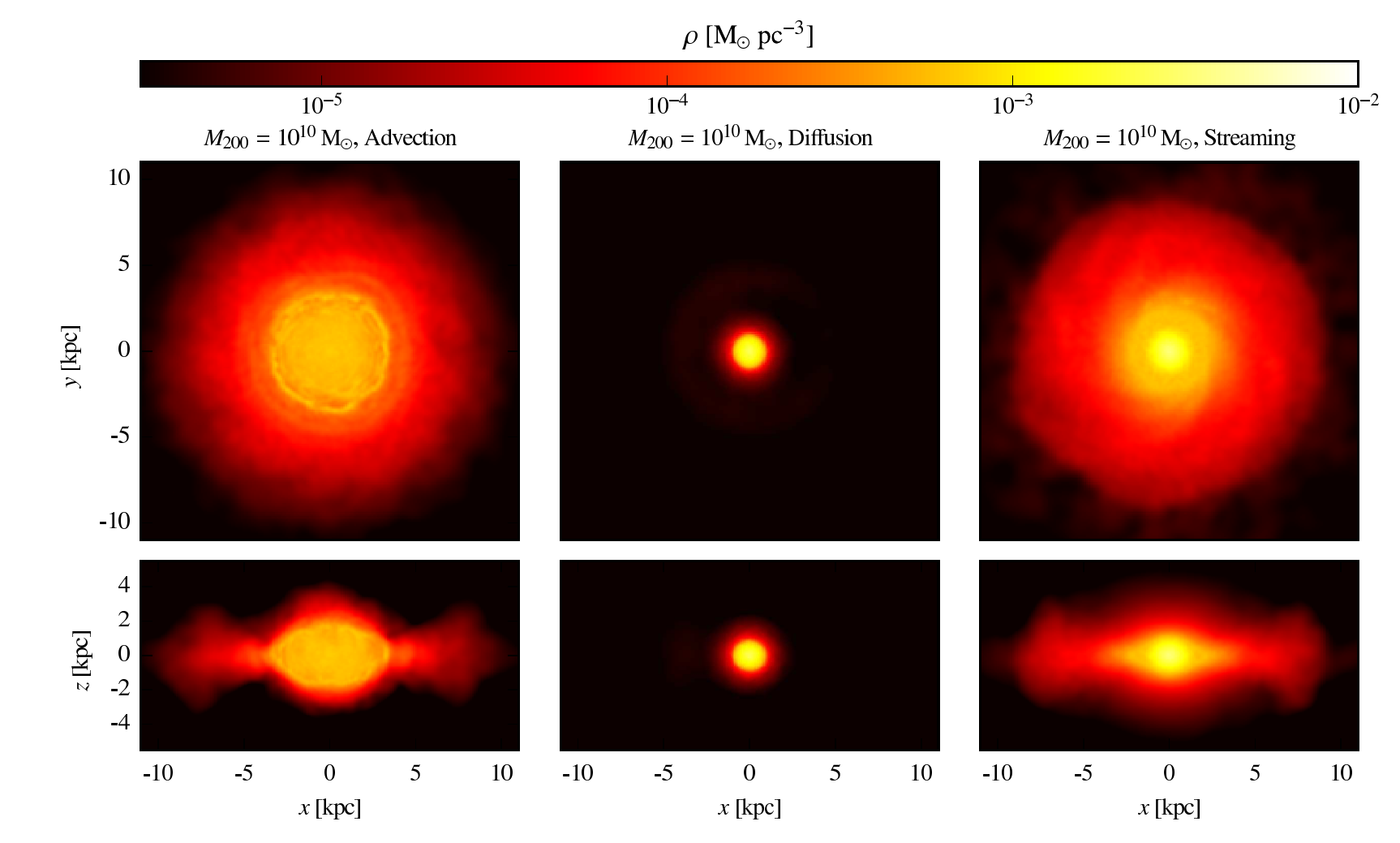}
\includegraphics{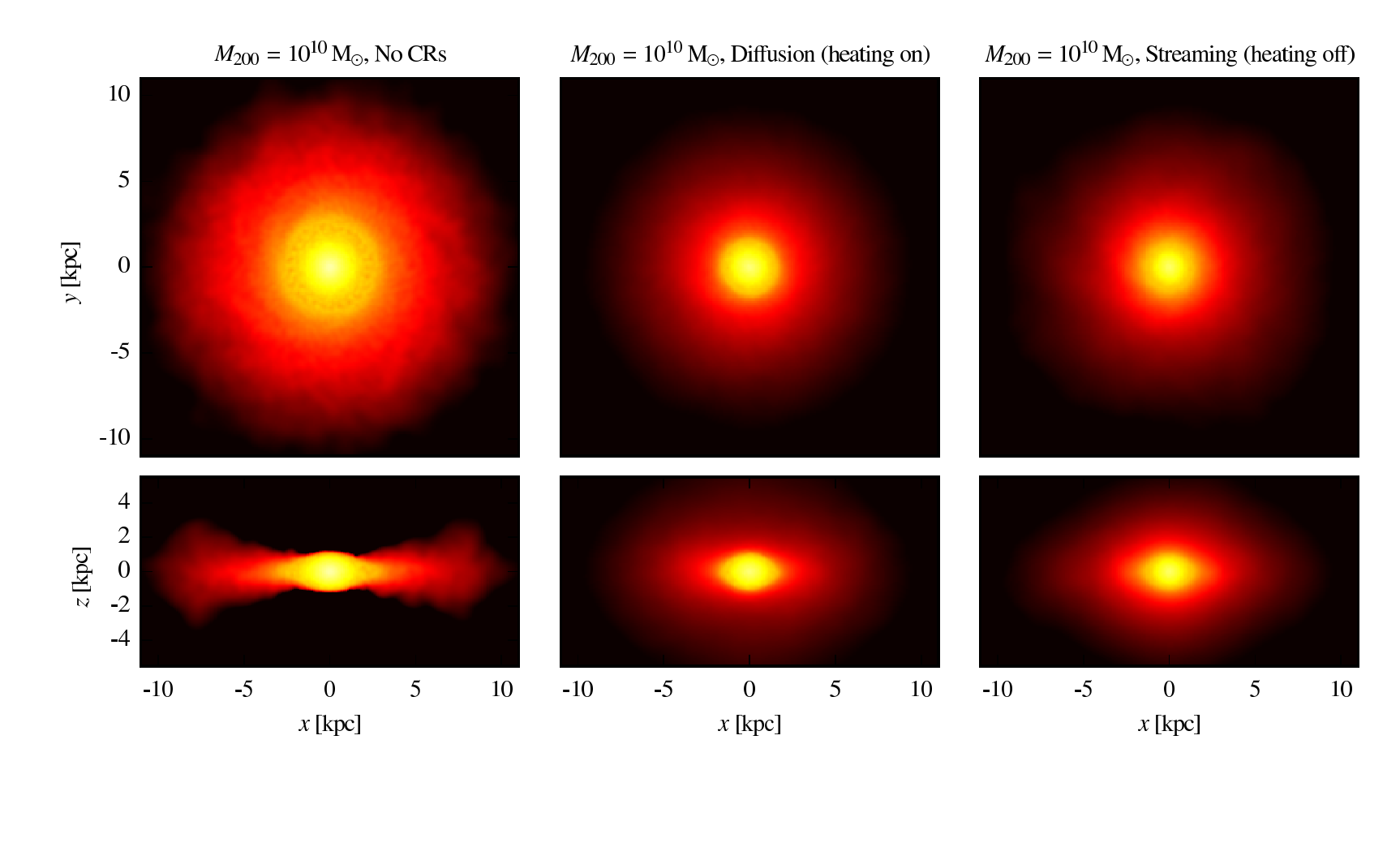}
\caption{Face-on and edge-on slices of mass density $\rho$ through the centre of the $10^{10}\ \rmn{M}_\odot$ halo. Shown are the six different simulation models after 1 Gyr. The strongest outflows of the CR diffusion model have dramatic impact on gas morphology and remove the disk (top middle panels). In contrast, the Alfv{\'e}n wave losses render the CR streaming model  less efficient in driving a powerful outflow from the disk and thus leave its morphology intact (top right panels). Interestingly, the two strawman models (diffusion with heating on and streaming with heating off) exhibit rather similar density morphologies.}
\label{rho10}
\end{figure*}

\begin{figure*}
\includegraphics{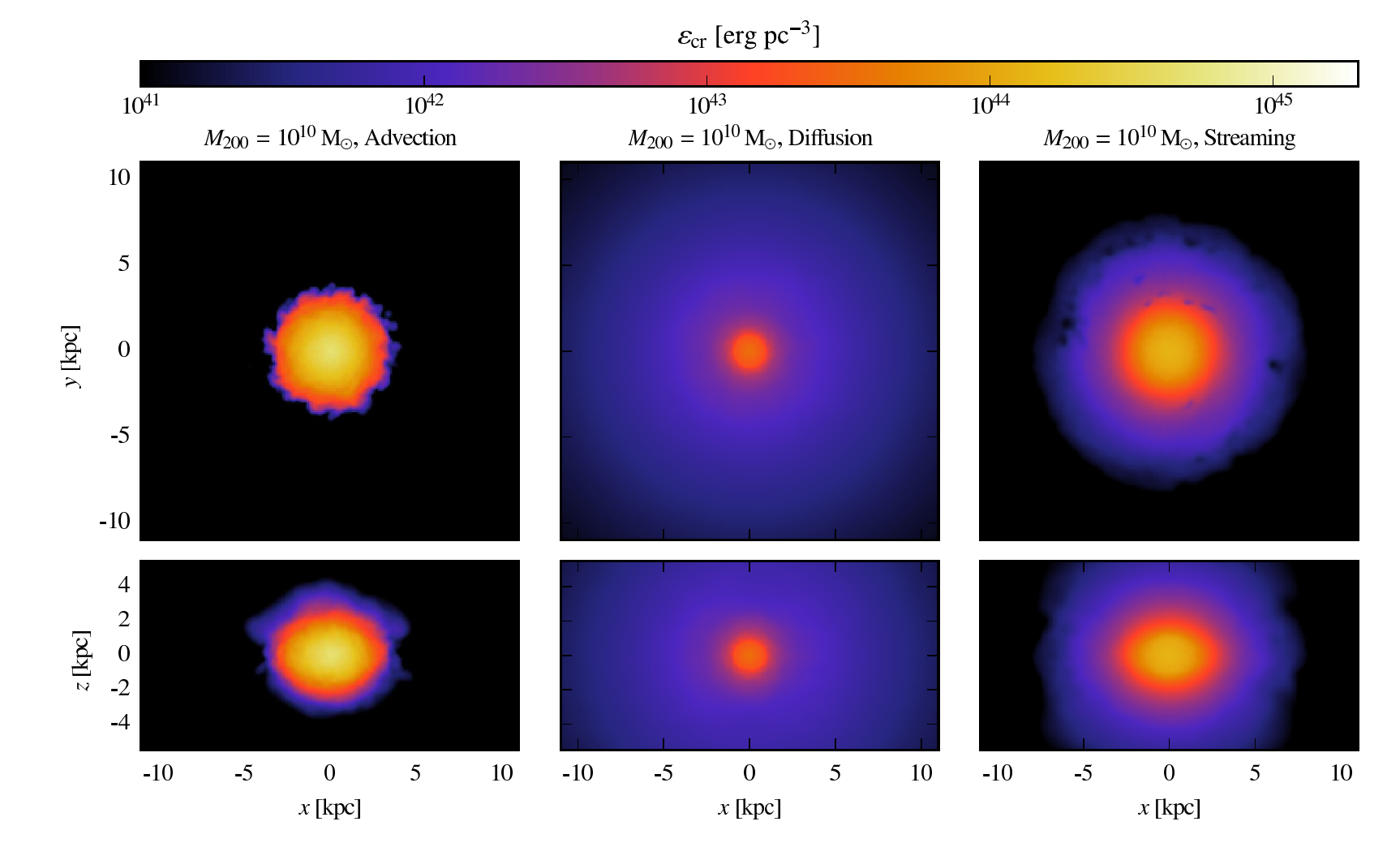}
\includegraphics{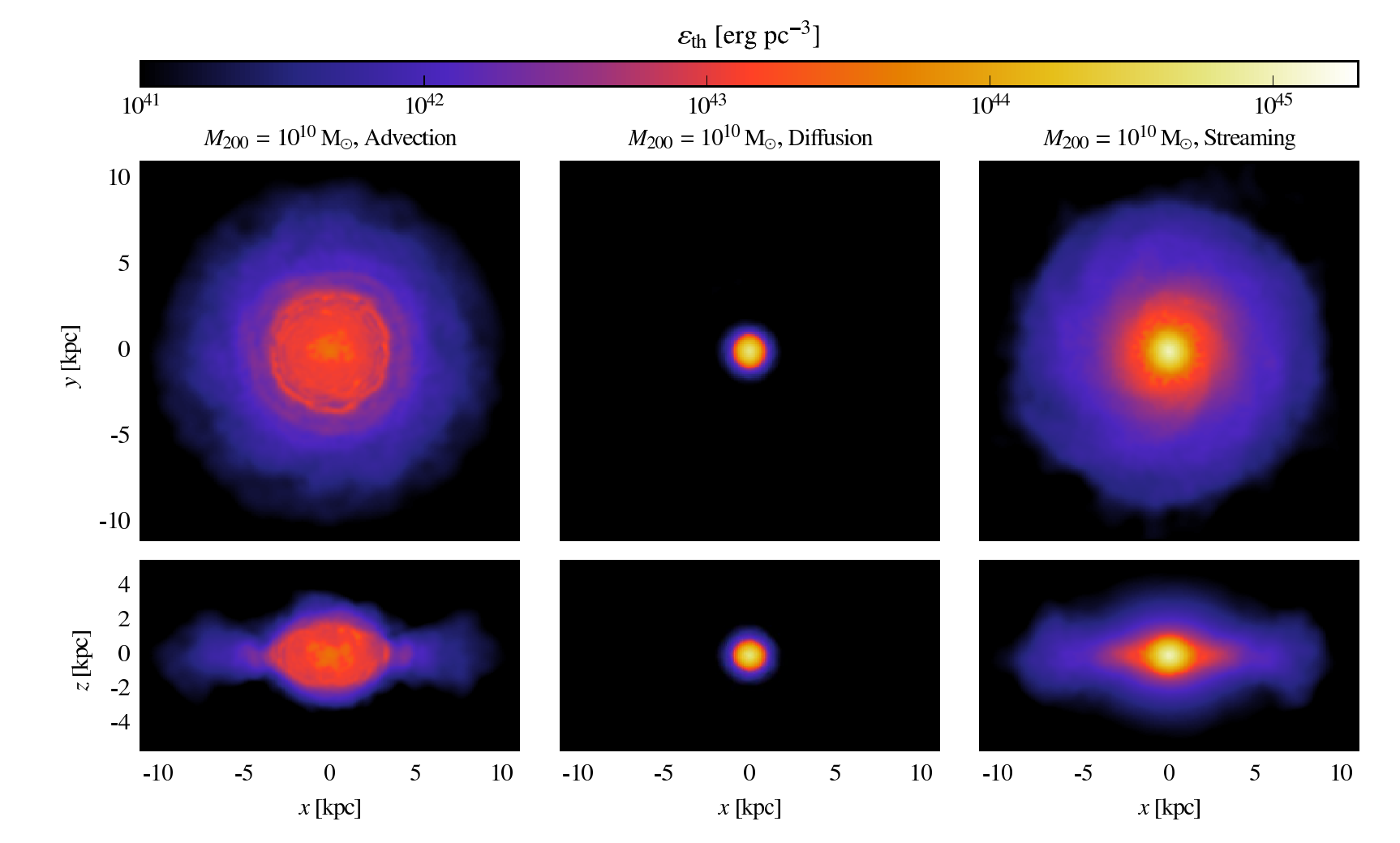}
\caption{Face-on and edge-on slices of CR energy density $\varepsilon_\mathrm{cr}$ (top six panels) and thermal energy density $\varepsilon_\mathrm{th}$ (bottom six panels) through the centre of the $10^{10}\ \rmn{M}_\odot$ halo after 1 Gyr, using the same colour scale. In the CR advection model (left), the CRs are tied to the gas disk and dominate the pressure at 1 Gyr. In contrast, the isotropic CR diffusion model (middle) exhibits a narrowly peaked CR distribution that drove an outflow, leaving behind a compact dwarf spheroidal morphology in $\varepsilon_\mathrm{th}$. The morphologies of $\varepsilon_\mathrm{cr}$ and $\varepsilon_\mathrm{th}$ in the streaming model (right) appear as a superposition of the other two models.}
\label{ecr10}
\end{figure*}

\begin{figure*}
\includegraphics{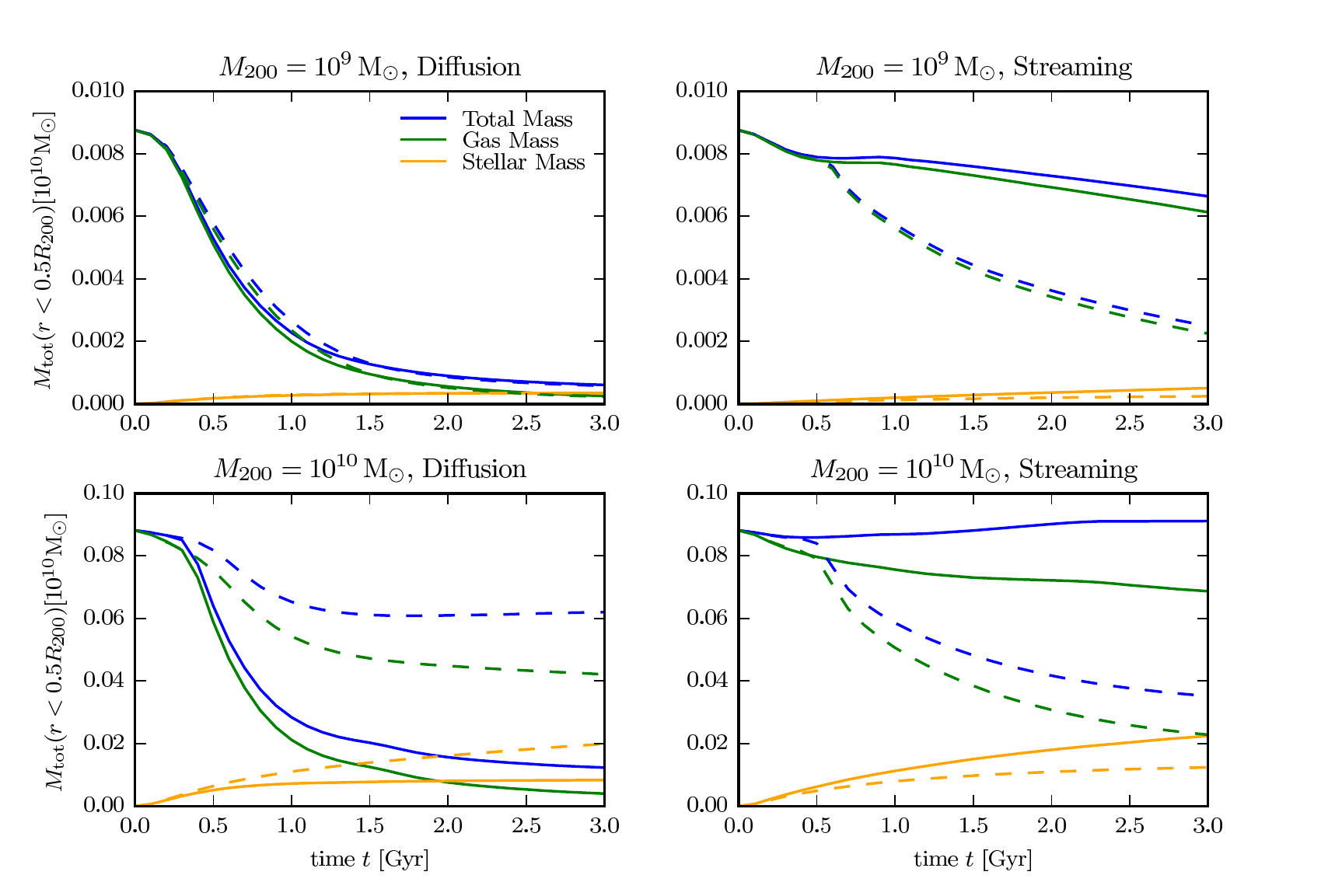}
\caption{Mass within central $0.5R_\textrm{200}$ of the 10$^{9}$ (top) and 10$^{10}\ \rmn{M}_\odot$ (bottom) halos, broken down into gas (green), stellar (yellow), and total mass (blue). Left: CR diffusion models as described (solid) and modified (dashed) to include the Alfv{\'e}n wave cooling term $\vA\bm{\cdot\nabla} P_\CR$. Right: CR streaming model as described (solid) and modified (dashed) to drop the Alfv{\'e}n wave cooling term.}
\label{masslosses}
\end{figure*}

\section{Simulation results}
\label{sec:results} 

\subsection{Isolated galaxy formation}\label{sec:galaxies}

We initialise our isolated galaxy simulations with a fixed Navarro-Frenk-White \citeyearpar[NFW,][]{navarro97} dark matter potential with a total halo mass of $M_{200} = 10^9$ and $10^{10}\, \rmn{M}_\odot$, respectively. The choice of these lower halo masses is motivated by the finding in \citet{uhlig12} that feedback due to CR streaming is more effective in these shallow potential wells. Our goal is to compare and contrast different assumptions for CR transport, rather than to do a detailed parameter study. The corresponding virial radii are $R_{200}=16.2$ and $34.9$ kpc.  The density profile is characterised by a concentration parameter, which we keep fixed at a value of $c_{200}=12$ for our two halos. Hence, these are scaled versions of each other which would evolve in a self-similar way if we only considered gravity and hydrodynamics. We assume that the gas is initially in hydrostatic equilibrium and adopt the universal gas fraction of 0.155. The gas is assumed to follow the same NFW profile, and to rotate slowly with a dimensionless spin parameter of $\lambda_\rmn{spin}=0.05$ and with a radial distribution of specific angular momentum in agreement with results from cosmological simulations \citep{bullock01}. Also, note that the halo gas profile is no longer in hydrostatic equilibrium once we put in rotation. Thus, early on, there is a small amount of mass loss, though eventually the halo gas settles to a new equilibrium profile. We use $10^5$ SPH particles to represent the gas distribution and 64 smoothing neighbours in all our simulations.

At the onset of the simulation the gas starts to cool radiatively, loses thermal pressure support, conserves it specific angular momentum and collapses to form a rotationally supported galactic disc in the centre of our halo. Once the gas exceeds a critical density threshold, a fraction of the gas is turned into stars. Depending on the type of simulation, supernova remnants accelerate CRs and the simulation follows their transport. We run a total of 6 CR models for each halo mass, as described in Section~\ref{sec:sim_overview} and summarised in Table~\ref{tab:models}.

Figure~\ref{SFRplot} shows SFRs (left panels) and mass loss relative to simulations without CRs (right panels) as a function of time in the first four models. Clearly, CR feedback always reduces the star formation considerably with a suppression factor that increases towards smaller galaxies. However, the star formation histories depend critically on the specific CR physics considered. While CR diffusion drives a strong outflow in both halos and reduces the amount of gas available for star formation, purely advected CRs are tied to the gas disk and provide additional pressure support that moves the gas to lower densities and thus decreases the amount of {\em dense} gas necessary for star formation. The CR streaming model generates weaker outflows and hence suppresses star formation by a combination of both effects.

Figure~\ref{ecr_bar} shows the average CR energy density
$\langle\eps_\rmn{cr}\rangle$ within a cylindrical disk of thickness $|z|<z_\rmn{disk}$ and radius in the disk plane $r<r_{\rmn{disk}}$. For the $10^9\ \msun$ halos we use $z_\rmn{disk}=1$~kpc and $r_\rmn{disk}=5$~kpc and double these values for the $10^{10}\ \msun$ halos.  Initially, the CR energy density rises steeply and then levels off in our CR advection model. While $\langle\eps_\rmn{cr}\rangle$ acquires a constant value in the $10^9\ \msun$ halo, the increased density in the $10^{10}\ \msun$ halo causes a noticeable decrease of CR energy at late times as a result of Coulomb and hadronic CR cooling. In the active CR transport models, CRs stream/diffuse away from their injection sites, thus reducing the average CR energy density in the disk. In comparison to our CR streaming model, CR diffusion drives a strong outflow that additionally advects CR energy into the halo, implying a dramatic reduction of $\langle\eps_\rmn{cr}\rangle$.

In addition to our physical CR models, in Fig.~\ref{ecr_bar} we also show CR models with the wave heating term turned on for diffusion and artificially turned off for streaming.  Turning the wave heating exchange off for CR streaming (Fig.~\ref{ecr_bar}, red dashed) allows the CRs to keep most of their energy at early times so that more CRs accumulate in the disk and are available for driving a more powerful wind in comparison to the CR streaming case that carries gas and CRs into the halo. Conversely, the effect of turning on the wave heating exchange for CR diffusion (Fig.~\ref{ecr_bar}, orange dashed) is to take a significant amount of energy out of the CRs at early times. This reduces the CR pressure gradient and causes a weaker breeze in comparison to the strong CR-diffusion driven wind. As a result, there are less CRs lost to the breeze and $\langle\eps_\rmn{cr}\rangle$ remains larger in comparison to the pure diffusion model at late times in the $10^{10}\ \msun$
halo.

We dissect the physical reasons for these particular behaviors by displaying the vertical velocity ($v_z$) in edge-on slices, the gas density ($\rho$), the CR and thermal energy density ($\eps_\rmn{cr}$ and $\eps_\rmn{th}$) for the 10$^{10}\ \rmn{M}_\odot$ halo models in Figs.~\ref{vel10}, \ref{rho10}, and \ref{ecr10}; all show slices through the centre of the halo at a time of 1 Gyr after the start of the simulation. These visual impressions are complemented by the time evolution of the total mass contained within the central $0.5R_\textrm{200}$ of the 10$^{9}$ and 10$^{10}\ \rmn{M}_\odot$ halos in Fig.~\ref{masslosses}.

Consider first the simulation without CRs. By 1 Gyr the gas has formed a disk, and gas continues to fall onto the disk, continually forming stars at a large rate. This behavior is notably changed for the case of pure advective CR transport. The CRs that are injected into the ISM cannot escape the disk since they cannot stream or diffuse relative to the gas, and they do not have sufficient pressure dominance in the disk to launch a wind. Generally, a CR driven wind is only possible if CRs can stream or diffuse out to the halo, where they are energetically dominant \citep{uhlig12, salem14a}. They thus act as an additional source of pressure support for the disk, puffing it up (compare the leftmost panels of Fig.~\ref{rho10}). Gas still falls onto the disk, but because the disk itself is less dense, the average star formation rate is much lower (Fig.~\ref{SFRplot}). However, as soon as the CRs cool by interacting hadronically and through Coulomb collisions with the ISM, collapse can start again and the SFR increases. Additionally, CR-driven buoyancy instabilities cause the CR over-pressured gas to rise from the disk potential, thereby inducing an increased small-scale velocity field in the disk in comparison to the simulation without CRs (cf.{\ }the leftmost panels of Fig.~\ref{vel10}).  As a result, the SFR shows an oscillating behavior \citep[Fig.~\ref{SFRplot}, see also][]{jubelgas08}.

Now consider the case with CR streaming. As CRs stream away from their acceleration sites, they are able to leave the disk. As a result the oscillations in the SFRs vanish and allow the gas to reach higher densities again. This enhances the SFR in comparison to the advection-only model. Most importantly, the $\bm{\nabla}P_\CR$ force term in the momentum equation is capable of driving outflow of material (top right of Fig.~\ref{vel10}). However, these winds are launched at large heights above the disk, where the density is low, and so although the SFR is noticeably reduced, the mass loss from the disk is only somewhat larger than in the case without CRs (see bottom right of Fig.~\ref{masslosses}, solid lines).

The simulation with CR diffusion, in stark contrast, shows a strong wind launched much lower in the disk, driving a significant amount of mass loss from the halo (see top left of Fig.~\ref{vel10} and bottom left of Fig.~\ref{masslosses}, solid lines). This effect is so strong that it alters the shape of the galaxy (top middle of Fig.~\ref{rho10}), leaving a small dwarf spheroidal morphology behind and leading to the lowest overall SFR.

What is the reason for this dramatic difference of our CR streaming and diffusion simulations? One marked difference is the additional CR cooling term $|\vA\bm{\cdot\nabla} P_\CR|$ in the case of CR streaming. By contrast, the CR diffusion term conserves CR energy and CRs only suffer adiabatic losses across the rarefaction wave that they generate by driving gaseous outflows. In order to test this hypothesis we run two alternate models of galaxy formation and simulate CR streaming {\em without} the Alfv{\'e}n cooling term as well as CR diffusion {\em with} the Alfv{\'e}n cooling term included. The results are shown in the bottom panels of Figs. \ref{vel10}, \ref{rho10}, and \ref{ecr10}, and in figure \ref{masslosses} which displays the time evolution of the total central mass distribution. The solid lines correspond to the regular models and the dashed lines denote the alternate models with the modified Alfv{\'e}n cooling term.

Figure \ref{masslosses} makes the clear point that a halo with CR streaming but \emph{no} wave heating losses will eject much more mass than the same halo with wave heating losses. The same is true for the halos using diffusion---CRs with no losses to wave heating push out much more mass than CRs that suffer these heating losses. This comparison clearly demonstrates that Alfv{\'e}n wave cooling is in large parts responsible for the inefficiency of CR streaming driven winds in larger halos, at least for our equilibrium model that adopts for the streaming speed $v_\rmn{A}\approx c_\rmn{s}$ \citep{uhlig12}. Note from Table~\ref{tab:models} that the maximum halo velocities always exceed the vertical disk velocities, indicating that the CR-driven wind is accelerating as a result of the forcing applied by the CR pressure gradient.

The radiative cooling time of the average ISM in the disk ($kT\sim10^4$~K and $n\sim1\,\rmn{cm}^{-3}$) is orders of magnitude smaller than the Alfv{\'e}n or sound crossing time of the vertical scale height which reinforces the argument that any energy deposited by CR heating is quickly radiated away. However, this is not true in the hot, low density coronae of higher mass halos ($kT\sim10^6$~K and $n\sim10^{-2}\,\rmn{cm}^{-3}$) with a radiative cooling time of $\tau_\rmn{rad}\approx200$~Myr. In such an environment, the Alfv{\'e}n or sound crossing time across the thick disk is $\tau_\rmn{A}\sim 1\,\rmn{kpc}/(150\,\rmn{km~s}^{-1})\sim6$~Myr, and thus much smaller than $\tau_\rmn{rad}$ so that the deposited CR energy can be retained and used for an additional thermal driving of the wind.

We caution that outflow and star formation rates will ultimately depend on the diffusion coefficient and magnetic field strengths, which motivates more realistic magneto-hydrodynamical simulations to consolidate this finding. Moreover, anisotropic CR transport will modify the wind acceleration by tying the CRs for longer time scales to the predominantly toroidal magnetic field of galaxies that has been amplified by the galactic shear \citep{Pakmor2016}. However, we can state that for the adopted (simple) assumptions, properties of galactic winds change considerably when CRs evolve via streaming instead of diffusion.

However, Fig.~\ref{masslosses} also shows that there are remaining differences. In order to understand those and how exactly CR streaming differs from diffusion in their impact on the hydrodynamics, we will turn our attention to simplified numerical experiments of these CR transport processes.

\begin{figure*}
\includegraphics[trim=3cm 0cm 0cm 0cm]{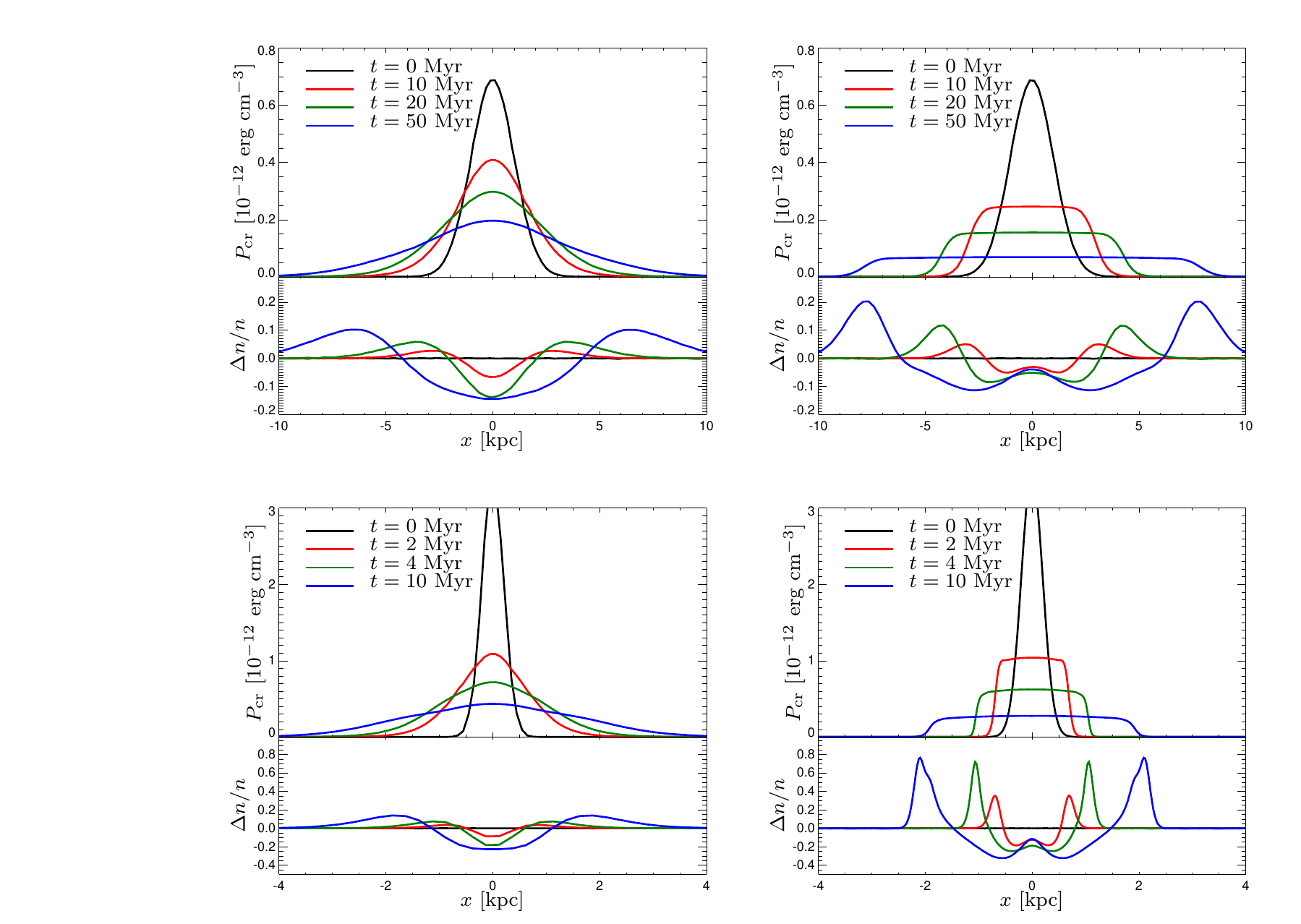}
\caption{CR pressure and density profiles for the homogeneous background diffusion (left) and streaming (right) simulations, starting with broad (top) and narrow (bottom) initial CR peaks. Note the different scales on the x-axis for the top and bottom plots.}\label{HBGPlots}
\end{figure*}

\subsection{Homogeneous flux tube models} \label{sec:fluxtubes}

We compare the dynamics of streaming and diffusing CRs for two different initial CR profiles in a homogeneous flux tube model. The background gas density in these simulations is set to $n=0.012$ cm$^{-3}$. The system evolves adiabatically, i.e. we ignore radiative cooling. The temperature and magnetic field are chosen such that the adiabatic sound speed $c_\rmn{s}$ and the Alfv\'en speed $v_\rmn{A}$ are both 100~km~s$^{-1}$ at this density. The domain is a long periodic box of 100~kpc length, with a 10~kpc by 10~kpc cross section\footnote{An exception is the sharp-peaked initial distribution with streaming. The required resolution here was very high, so we only simulate a 10 kpc long box.}. The lowest resolution runs are filled with $10^4$ SPH particles with glass initial conditions\footnote{A glass-like distribution is obtained by randomly sampling the particle positions and then evolving the hydrodynamic equations for this distribution until reaching approximate hydrostatic equilibrium.} of mean inter-particle spacing of 1 kpc. Other runs have resolution increased as necessary: the broad initial Gaussian evolved with streaming uses $8\times 10^4$ particles (mean particle spacing 0.5 kpc). The sharp initial Gaussian with diffusion uses $27\times 10^4$ particles (mean particle spacing 0.33 kpc). The sharp initial Gaussian with streaming uses $4096\times 10^4$ particles (mean particle spacing 0.0625 kpc).

The first setup has a plane-parallel ``broad'' Gaussian of width 1 kpc for the initial CR pressure profile with $X_{\CR}\equiv P_\CR/P_\gas=1$ at the central maximum.  This length scale $L$ is chosen such that the initial diffusive energy flux $\kappa \nabla P_\CR\sim\kappa P_\CR / L$ in the diffusion-only runs is roughly the same as the streaming energy flux $v_\rmn{A} P_\CR$ in the streaming-only runs (the diffusion coefficient is $\kappa=3\ee{28}$~cm$^2$~s$^{-1}$).

We show the evolving CR profiles as CRs are diffusing in Fig.~\ref{HBGPlots}, top left. The CR pressure profile broadens with time while remaining nearly a Gaussian distribution. Small overdensities are generated and the gas contained in those is accelerated initially by the CR pressure gradient -- these then travel outward at the sound speed and cause rarefaction at the centre and an associated adiabatic cooling of the CRs.

In contrast, the CR profile flattens from the centre out for the constant streaming speed case (Fig.~\ref{HBGPlots}, top right). In this case CR wave heating is transferring a significant amount of energy to the gas at the positions of a steep CR pressure gradient. Energy conservation during this transfer ($\Delta E_\CR=\Delta E_\gas$) increases the total pressure of the composite fluid due to the harder equation of state of the thermal plasma.  This can be seen by evaluating the resulting gas pressure (of an initially cold fluid),
\begin{equation}
  P_\gas = \frac{\gamma_\gas-1}{\gamma_\CR-1} P_\CR \approx 2 P_\CR,
\end{equation}
assuming an ultra-relativistic CR population with $\gamma_\CR=4/3$ which dissipated all its energy. The resulting pressure force is also doubled and causes a density perturbation that is twice that of the diffusive case despite the matching CR fluxes in both cases (cf.\ top panels of Fig.~\ref{HBGPlots}).

This difference is magnified when we start with a sharper CR profile. A second setup uses an initial Gaussian CR pressure profile of width 0.2 kpc and $X_{\CR}=5$ at the central maximum, such that the total initial CR energy is the same as the first setup. The diffusive case is nearly the same as before (Fig.~\ref{HBGPlots}, bottom left): the CR pressure profile evolves almost exactly as a Gaussian, so starting with a sharper peak amounts to merely a slight offset in time. However, because the CR pressure gradients are much stronger here, the streaming case is considerably different because of the different CR fluxes (Fig.~\ref{HBGPlots}, bottom right). Most of the energy is transferred to the gas on a very short time scale (note the different time stamps in this plot) resulting in a substantial density spike.

We note that it is this property that precludes the (effective approximate) modelling of CR streaming with diffusion using a single constant value of the diffusion coefficient $\kappa$. Choosing $\kappa$ that is tailored to one situation such as CR transport of the equilibrium regime of a stratified CR pressure disk -- similar to our Galaxy today -- we obtain a diffusion velocity $v_\rmn{diff}\sim \kappa/L \sim 30\,\rmn{km\,s}^{-1}$ for $\kappa\approx3\ee{28}$ cm$^2$ s$^{-1}$ and $L\approx3$~kpc. If the CRs are however injected locally at a supernova remnant with $L\approx30$~pc, the resulting diffusion velocity would be 100 times larger and imply super-Alfv{\'e}nic streaming by the same factor, in conflict with the theory of CR transport as laid out in Sect.~\ref{sec:theory}. The assumption of a spatially constant diffusion coefficient is clearly over-simplistic.  Close to the sources, for a distance of the order of the coherence length of the large scale magnetic field, the CR density is dominated by freshly injected CRs from the source and the CR gradient can be large enough to strongly amplify the magnetic turbulence (e.g., through the non-resonant hybrid instability, \citealt{bell04}), thereby reducing the diffusion coefficient \citep[e.g.,][]{malkov13,nava16}.

\section{Conclusions}
\label{sec:conclusions} 

CRs are an attractive agent for driving galactic winds, but there has not been a consensus about the dominant CR transport mechanism: is it streaming or diffusion? Here we demonstrate that both mechanisms obey a completely different mode of propagation with significantly different dynamical evolution. This can be most easily summarised by comparing the advective streaming velocity $\vs$ with the dispersive diffusive transport velocity $\vdi$ of a CR population (shown in the isotropic approximation for simplicity):
\begin{equation}
  \vs=\bm{n}\,v_\rmn{A}
  \quad\mbox{and}\quad
  \vdi=-\kappa\frac{\bm{\nabla} \eps_{\rmn{c}}}{\eps_{\rmn{c}}}
  =\bm{n}\,\frac{\kappa}{L_\rmn{c}(\vec{x},t)},
\end{equation}
where $\bm{n}$ is a unit vector pointing down the CR gradient. Thus, pressure gradients are preserved by CR streaming, but not diffusion.  Due to the evolving CR gradient length $L_\rmn{c}(\vec{x},t)$ as a result of CR transport, it is impossible to match the streaming and diffusion fluxes at all times with a constant diffusion coefficient (as it is currently assumed in all calculations of galaxy formation with CR hydrodynamics). As a result, diffusion with a constant coefficient can lead to unphysically high CR fluxes, particularly early on when CR profiles are sharply peaked.

There is a second major difference: the excitation of Alfv{\'e}n waves is associated with an energy loss term for the CR population, which appears as an energy gain term for the thermal energy.  While this should double the generated thermal pressure for each unit of dissipated CR pressure, the energy is quickly radiated away if it is deposited into the dense phase of the interstellar medium, leaving little dynamical impact on the gas. Wave excitation due to streaming is thus a net loss mechanism for CRs in gas where the cooling time is short. By contrast, the frequent assumption of CR diffusion while ignoring wave excitation allows the CRs to evolve adiabatically, and overestimates the dynamical effect of CRs. It is inconsistent to ignore wave excitation in modelling CR winds, since it is only via these waves that CRs can couple and transfer momentum to the gas. In our work, we have found the energy losses due to Alfv{\'e}n wave heating to be the dominant difference between models in which either streaming or diffusion dominates.

To demonstrate this difference, we simulated a series of simplified models of isolated galactic halos that form via gravitational collapse.  We find that CR feedback always regulates the star formation considerably and has a suppression factor that increases towards smaller galaxies. However, the regulating mechanism depends critically on the type of CR physics considered.  CR diffusion drives strong outflows as CRs diffuse up from the disk and smoothly accelerate the gas by the CR pressure gradient. While these diffusively-driven winds reduce the amount of gas available for star formation, purely advected CRs are tied to the gas disk and provide additional non-thermal pressure support that moves the gas to lower densities and thus decreases the amount of dense gas necessary for star formation. The CR streaming model generates weaker outflows and hence suppresses star formation by a combination of both effects. These weaker outflows are mainly due to two effects. (i) While CR diffusion-driven outflows start to dominate the pressure balance already deep in the potential and can thus accelerate a larger column of gas, CR streaming-driven winds start higher up in the atmosphere and can only accelerate a smaller gas column. (ii) CR Alfv\'en wave losses dissipate a substantial CR energy already in the disk, thereby reducing the available CR pressure for the wind driving. We note that our results are in full agreement with earlier works on CR streaming-driven winds \citep{uhlig12} or CR diffusion-driven winds \citep{booth13}.

These conclusions come with a number of caveats. (i) In this exploratory work, we only modelled isotropic CR diffusion and streaming in order not to confuse magnetic effects with those resulting from CR transport. Following the (more correct) anisotropic transport in MHD simulations of forming galaxies, it was shown that wind properties (after the onset of the outflow) are similar for isotropic and anisotropic diffusion while there are significant differences in the large-scale magnetic dynamo \citep{Pakmor2016}. (ii) In order to accurately model CR streaming we need to follow magneto-hydrodynamics since the Alfv\'en speed as the characteristic CR streaming speed depends linearly on the field strength. Since galaxy formation is likely accompanied by a magnetic dynamo that amplifies the small seed field, this could argue in favour of an initial phase of CR feedback that more closely resembles our pure-advection model before the growing field strength allows for a sizeable CR streaming speed, which accelerates the winds (in the self-confinement picture). (iii) Finally an accurate model for CR diffusion requires a self-consistent diffusion coefficient that must go beyond the constant approximation. We have seen that the latter is a poor approximation in the self-confinement limit (linear damping mechanisms such as turbulent damping map onto the streaming term, while non-linear damping mechanisms have a different mathematical character; none of these exhibit classical diffusive behavior), though it may be plausible if small scale magnetic structure sets the mean free path.  Our presented results are encouraging in that CR feedback plays a major role in galaxy formation and hopefully motivates future work to address the caveats of the presented work.

\section*{Acknowledgements}
We thank Volker Springel, R{\"u}diger Pakmor, and Ellen Zweibel for helpful discussions. JW and SPO acknowledge NASA grants NNX12AG73G and NNX15AK81G for support. CP gratefully acknowledges support by the European Research Council under ERC-CoG grant CRAGSMAN-646955 and by the Klaus Tschira Foundation. JW acknowledges the University of Wisconsin-Madison and NSF grant AST 1616037 for support. SPO acknowledges the hospitality of KITP; this research was supported in part by the National Science Foundation under Grant No. NSF PHY11-25915.

\bibliography{master_references}

\label{lastpage}


\appendix

\section{Simulations without Hydrodynamics}
\label{sec:tests}

\begin{figure}
\includegraphics[trim=3cm 0cm 0cm 0cm]{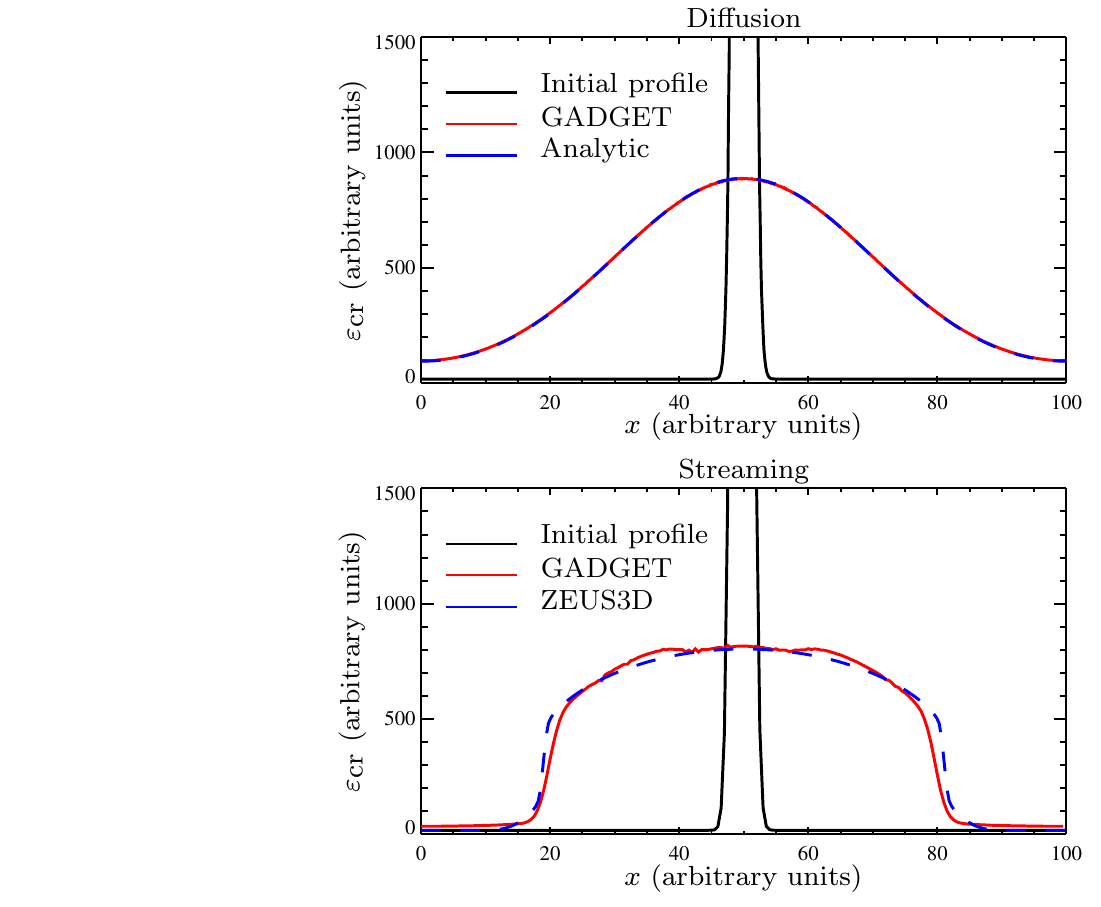}
\caption{Initial Gaussian distribution for the CR energy density $\varepsilon_\rmn{cr}$ is evolved with a constant diffusion coefficient (top) and with constant streaming speed (bottom). In the top panel, the simulation result is compared with the analytic solution (equation \eqref{periodicsol}) at time $t=2.0$. In the bottom panel, the GADGET results are compared with those of the ZEUS grid code in 1D at time $t=0.2$.}\label{nohydrotests}
\end{figure}

To perform a basic test of the explicit finite difference schemes of the CR diffusion \citep{jubelgas08} and streaming modules \citep{uhlig12} in GADGET-2, we ran simulations with a static background, i.e., without accounting for the CR pressure forcing on the thermal gas. These runs solely test the evolution of the CR quantities. We start with a uniform background gas and a Gaussian CR specific energy, with CR spectral index $\alpha=2.5$ and low-momentum cutoff of the CR distribution (in units of $m_\rmn{p} c$) $q_0=10$. This setup is then evolved using (i) a constant diffusion coefficient, and (ii) a constant streaming speed.

For a constant diffusion coefficient, there exists an analytic solution. A Gaussian profile with initial width $\sigma_0$ undergoing diffusion with diffusion coefficient $\kappa$ evolves in time according to
\begin{equation}\label{analyticdiff}
\eps_\rmn{cr,1}(x,t)=\eps_\rmn{cr,max}\sqrt{\frac{\sigma_0^2}{\sigma_0^2+2\kappa t}}\exp\left(\frac{-x^2}{2(\sigma_0^2+2\kappa t)}\right).
\end{equation}
That is, an initial Gaussian profile evolving by pure diffusion remains a Gaussian with a width that widens with time. To account for the periodicity of the GADGET-2 simulation, we superimpose the above analytic solution with equivalent solutions offset by 100 in either direction:
\begin{equation}\label{periodicsol}
\eps_\rmn{cr,sol}(x,t)=\eps_\rmn{cr,1}(x,t)+\eps_\rmn{cr,1}(x+100,t)+\eps_\rmn{cr,1}(x-100,t)
\end{equation}
In principle we would add an infinite number of offset Gaussians, but the contributions past the immediate neighbouring peaks are negligible.

The top panel of Fig.~\ref{nohydrotests} shows that the simulation of CR diffusion provides an excellent match to the analytic solution (equation~\ref{analyticdiff}). The streaming case has no analytic solution, but we can compare it to other numerical codes. The bottom plot of Fig.~\ref{nohydrotests} shows the streaming result, compared with the same setup evolved in a modified version of the ZEUS code \cite[see][]{Guo08a}. Again, the results match well, although the GADGET-2 result is more diffusive. This behavior is expected in an SPH code, and was in fact seen in \cite{uhlig12}.

\section{Resolution Tests}
\label{sec:res}

\begin{figure*}
\includegraphics{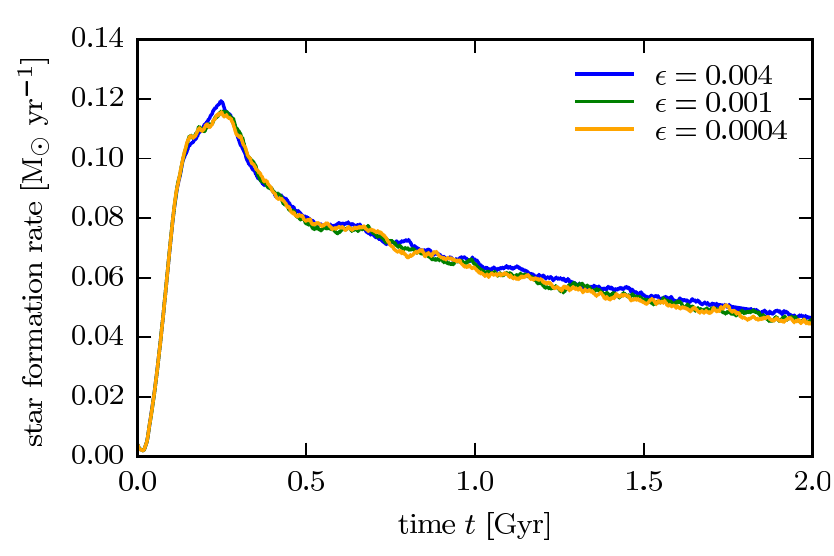}
\includegraphics{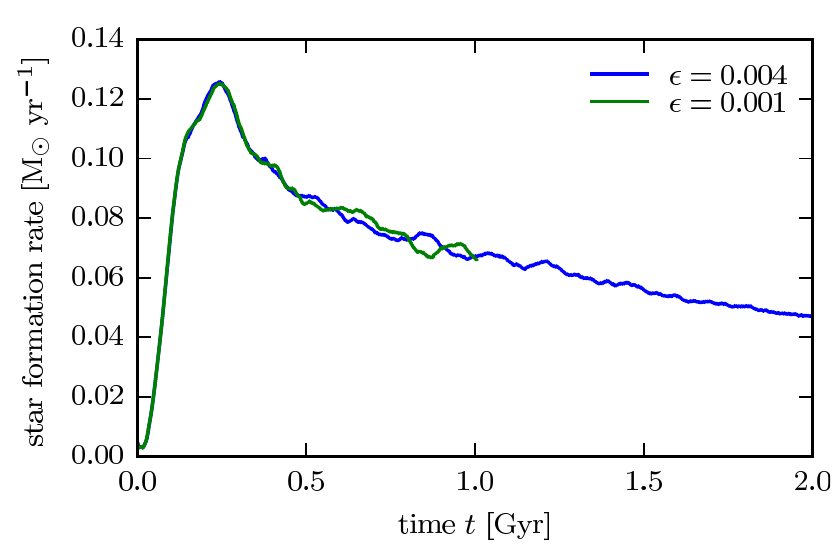}
\caption{Time resolution tests for a simulation of the 10$^{10}\ \rmn{M}_\odot$ with CR streaming. Left: 10$^5$ SPH particle runs with progressively more conservative choices for the Courant streaming factor $\epsilon$. Right: 10$^6$ SPH particle runs. Note that the halos used in these resolution tests have a concentration parameter of $c_{200}=7.2$, as opposed to $c_{200}=12$ used in the main body of this paper. We only run the simulation with $\epsilon=0.001$ and 10$^6$ particles for 1~Gyr to demonstrate convergence.}
\label{timetests}
\end{figure*}

\begin{figure*}
\includegraphics{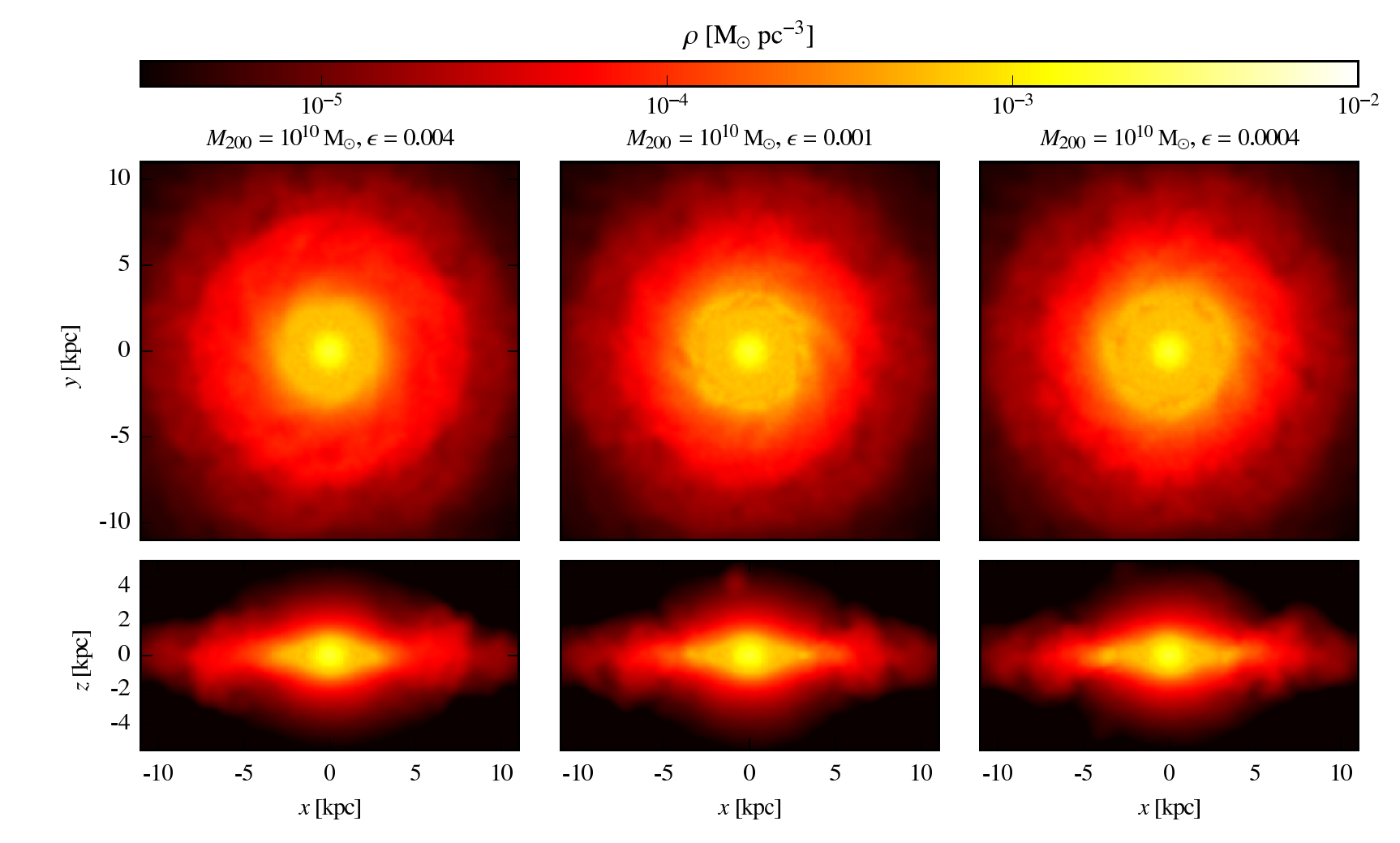}
\caption{Face-on and edge-on slices of mass density $\rho$ through the $10^{10}\ \rmn{M}_\odot$ halo of the CR streaming model at 1 Gyr with $10^{5}$ particles. Shown are results for progressively more conservative choices of the Courant streaming factor $\epsilon$. Note that the halos used in these resolution tests have a concentration parameter of $c_{200}=7.2$, as opposed to $c_{200}=12$ used in the main body of this paper.}
\label{rhores}
\end{figure*}

\begin{figure*}
\includegraphics{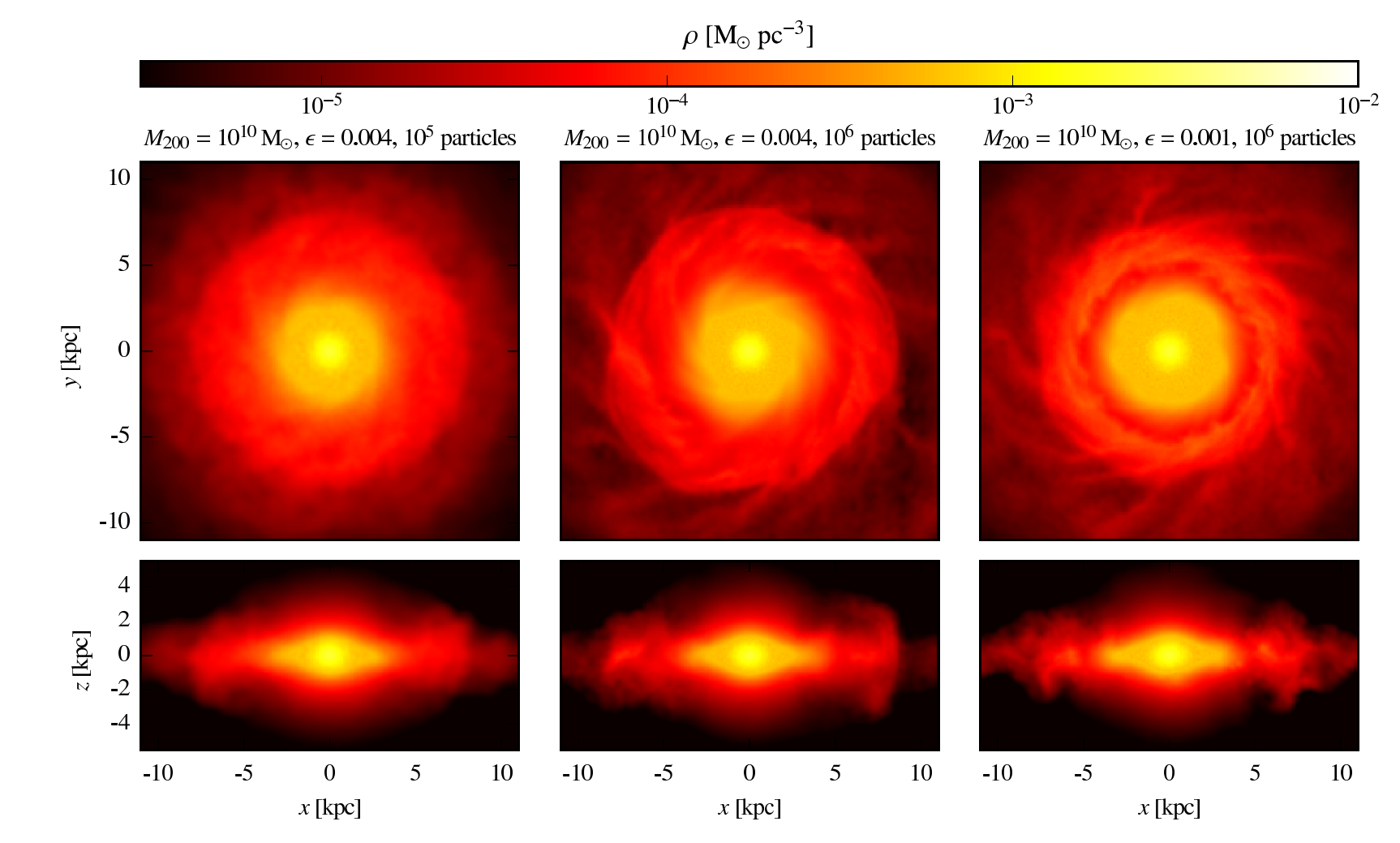}
\caption{Face-on and edge-on slices of mass density $\rho$ through the $10^{10}\ \rmn{M}_\odot$ halo of the CR streaming model at 1.0 Gyr. Left: we show the fiducial run with 10$^5$ SPH particles and $\epsilon=.004$. The middle and right panels show simulations with 10$^6$ SPH particles. Note that the halos used in these resolution tests have a concentration parameter of $c_{200}=7.2$, as opposed to $c_{200}=12$ used in the main body of this paper.}
\label{rhores_1M}
\end{figure*}

Here we provide a selection of resolution tests for the adopted choices of numerical parameters associated with the explicit finite difference scheme of the CR streaming module in GADGET-2.  Figure~\ref{timetests} shows time resolution tests with progressively more conservative choices for the Courant streaming factor $\epsilon$. Figures \ref{rhores} and \ref{rhores_1M} show face-on and edge-on slices of mass density $\rho$ through the $10^{10}\ \rmn{M}_\odot$ halo of the CR streaming model at 1 Gyr. We vary again the Courant streaming factor $\epsilon=\{0.004,0.001,0.0004\}$ and the number of SPH particles $N=\{10^5,10^6\}$. We see that the simulation with $N=10^6$ particles has the tendency to somewhat underestimate the density distribution in the outer disk, indicating the need for a more conservative choice for Courant streaming factor of $\epsilon\leq0.001$.  However, these convergence tests demonstrate that our choice of $\epsilon=0.004$ and $N=10^5$ produces numerically converged results.

\end{document}